\documentclass[pre,superscriptaddress,showpacs,preprint]{revtex4}
\usepackage{amsmath,graphicx,amssymb,amsfonts}

\begin{document}\title{Granular Solid Hydrodynamics}
\author{Yimin Jiang}
\affiliation{Theoretische Physik, Universit\"{a}t
T\"{u}bingen,72076 T\"{u}bingen, Germany}

\affiliation{Central South University, Changsha 410083,
China}

\author{Mario Liu}

\affiliation{Theoretische Physik, Universit\"{a}t
T\"{u}bingen,72076 T\"{u}bingen, Germany}

\date{\today}

\begin{abstract}
{\em Granular elasticity}, an elasticity theory useful
for calculating static stress distribution in granular
media, is generalized to the dynamic case by including
the plastic contribution of the strain. A complete
hydrodynamic theory is derived based on the hypothesis
that granular medium turns transiently elastic when
deformed. This theory includes both the true and the
granular  temperatures, and employs a free energy
expression that encapsulates a full jamming phase
diagram, in the space spanned by pressure, shear stress,
density and granular temperature. For the special case of
stationary granular temperatures, the derived
hydrodynamic theory reduces to {\em hypoplasticity}, a
state-of-the-art engineering model.

\end{abstract}  \maketitle

\tableofcontents

\section{Introduction\label{intro}}
Widespread interests in granular media were aroused among
physicists a decade ago, stimulated in large part by
review articles revealing the intriguing and improbable
fact that something as familiar as sand is still rather
poorly understood~\cite{1996-1,1996-2,1996-3,1996-4}. The
resultant collective efforts have since greatly enhanced
our understanding of granular media, though the majority
of theoretic considerations have focused either on the
limit of highly excited gaseous
state~\cite{Haff-1,Haff-2,Haff-3,Haff-4,Haff-5}, or that
of the fluid-like flow~\cite{chute-1,chute-2,chute-3}.
Except in some noteworthy and insightful
simulations~\cite{hh-1,hh-2,hh-3}, the quasi-static,
elasto-plastic motion of dense granular media -- of
technical relevance and hence a reign of engineers --
received less attention among physicists.

This choice is due at least in part to the highly
confusing state of engineering theories, where
innumerable continuum mechanical models compete,
employing vastly different expressions. Although the
better ones achieve considerable realism when confined to
the effects they were constructed for, these differential
equations are more a rendition of complex empirical data,
less a reflection of the underlying physics. In a
forthcoming book on soil mechanics by Gudehus, phrases
such as {\em morass of equations} and {\em jungle of
data} are deemed apt metaphors.

Most engineering theories are {\em
elasto-plastic}~\cite{elaPla-1,elaPla-2,elaPla-3}, though
there are also {\em hypoplastic}
ones~\cite{Kolym-1,Kolym-2}, which manage to retain the
realism while being simpler and more explicit. Both
adhere to the continuum mechanical formalism laid down by
Truesdell~\cite{trues-1,trues-2} who, starting from
momentum conservation, focuses on the total stress
$\sigma_{ij}$, and considers its dependence on the
variables: strain $\varepsilon_{ij}$, velocity gradient
$\nabla_jv_i$ and mass density $\rho$. Frequently, an
explicit expression for $\sigma_{ij}$ appears impossible,
incremental relations are then constructed, expressing
$\partial_t\sigma_{ij}$ in terms of $\sigma_{ij},
\nabla_iv_j, \rho$. Because the macroscopic energy (such
as its kinetic or elastic contribution) dissipates,
Truesdell does not include energy conservation in his
standard prescription.

In contrast, conservation of {\em total energy} is an
essential part of the {\em hydrodynamic approach} to
macroscopic field theories, pioneered in the context of
superfluid helium by Landau~\cite{LL6} and
Khalatnikov~\cite{Khal}. The total energy $w$ they
consider depends, in addition to the relevant macroscopic
variables such as $\rho$ and $v_i$, also on the entropy
density $s$. (There are different though equivalent ways
to understand $s$. The appropriate one here is to take it
as the summary variable for all implicit, microscopic
degrees of freedom. So the energy change associated with
$s$, always written as $(\partial w/\partial s){\rm
d}s\equiv T{\rm d}s$, is the increase of energy contained
in these degrees of freedom -- what we usually refer to
as heat increase.) When the macroscopic energy dissipates
into the microscopic degrees of freedom, the change in
entropy is such that the increase in heat is equal to the
loss of macroscopic energy, with the total energy $w$
being conserved.

The hydrodynamic approach~\cite{hydro-1,hydro-2} has
since been successfully employed to account for many
condensed systems, including liquid crystals
~\cite{liqCryst-1,liqCryst-2,liqCryst-3,liqCryst-4,liqCryst-5,liqCryst-6,liqCryst-7},
superfluid
$^3$He~\cite{he3-1,he3-2,he3-3,he3-4,he3-5,he3-6},
superconductors~\cite{SC-1,SC-2,SC-3}, macroscopic
electro-magnetism~\cite{hymax-1,hymax-2,hymax-3,hymax-4}
and
ferrofluids~\cite{FF-1,FF-2,FF-3,FF-4,FF-5,FF-6,FF-7,FF-8,FF-9}.
Transiently elastic media such as polymers are under
active consideration at
present~\cite{polymer-1,polymer-2,polymer-3,polymer-4}.

The main advantage of the hydrodynamic approach is its
stringency. In the Truesdell approach, apart from
objectivity, few general constraints exist for the
functional dependence of $\sigma_{ij}$ or
$\partial_t\sigma_{ij}$. Therefore, one needs to rely
entirely on experimental data input. In contrast, the
structure of the hydrodynamic theory is essentially given
once the set of variables is chosen. This is a result of
the constraints provided by energy conservation, which
enables one to fully determine the form of all fluxes,
including especially the stress $\sigma_{ij}$. These
expressions are given in terms of the energy's variables
and conjugate variables, they are valid irrespective what
form the energy $w$ has. (If $w$ is a function of $s,
\rho, \varepsilon_{ij}$, the conjugate variables are the
respective derivative: Temperature $T\equiv\partial
w/\partial s$, chemical potential $\mu\equiv\partial
w/\partial\rho$, and elastic stress
$\pi_{ij}\equiv-\partial w/\partial u_{ij}$.) We refer to
the fluxes as the structure of the theory, while taking
the explicit form of $w(s,\rho,u_{ij})$ as a scalar
material quantity.

There is little doubt that constructing a granular
hydrodynamic theory is both useful and possible: Useful,
because it should help to illuminate and order the
complex macroscopic behavior of granular solid; possible,
because total energy is conserved in granular media, as
it is in any other system. When comparing agitated sand
to molecular gas, it is frequently emphasized that the
kinetic energy, although conserved in the latter system,
is not in the former, because the grains collide
inelastically. This is undoubtedly true, but it does not
rule out the conservation of {\em total energy}, which
includes especially the heat in the grains, and in the
air (or liquid) between them.

To actually construct the granular hydrodynamic theory,
we need to start from some assumptions about the essence
of granular physics. Our choice is specified below,  and
argued for throughout this manuscript. As we shall see,
it is a guiding notion complete enough for the derivation
of a consistent hydrodynamic theory, the presentation of
which is the main purpose of the present manuscript. On
the other hand, we are fully aware that only future works
will show whether our assumptions are appropriate,
whether the resultant set of partial differential
equations is indeed ``granular hydrodynamics."

Granular motion may be divided into two parts, the
macroscopic one arising from the large-scaled, smooth
velocity of the medium, and the mesoscopic one from the
small-scaled, stochastic movements of the grains. The
first is as usual accounted for by the hydrodynamic
variable of velocity, the second we shall account for by
a scalar, the granular temperature $T_g$ -- although the
analogy to molecular motion is quite imperfect: The
grains do not typically have velocities with a Gaussian
distribution, and equipartition is usually violated. All
this, as we shall see, is quite irrelevant in the present
context.

$T_g$ may be created by external perturbations such as
tapping, or internally, by nonuniform macroscopic motion
such as shear -- as a result of both the grains will
jiggle and slide. Then the grains will loose contact with
one another briefly, during which their individual
deformation will partially relax. When the deformation is
being diminished, so will the associated static stress
be. This is the reason granular media can sustain static
stress only when at rest, but looses it gradually when
being tapped or sheared. And our assumption is, this
happens similarly no matter how the grains jiggle and
slide, and we may therefore parameterize their stochastic
motion as a scalar $T_g$. Our guiding notion is
therefore: Granular media are {\em transiently elastic};
the elastic stress relaxes toward zero, with a rate
$\tau^{-1}$ that grows with $T_g$, most simply as
$\tau^{-1}\sim T_g$.

In granular statics, the grains are at rest, hence
$T_g\equiv0$. With $\tau\sim T_g^{-1}$ infinite, granular
stress persists forever, displaying in essence elastic
behavior~\cite{J-L-1,J-L-2,J-L-3,ge-1,ge-2}. When
granular media are being sheared, because the grains move
nonuniformly and $T_g\not=0$, the stress relaxes
irreversibly. This is a qualitative change from the
elastic, purely reversible behavior of ideal solids. We
believe, and have some evidence, that it is this
irreversible relaxation that lies at the heart of plastic
granular flows. If true, this insight would greatly
simplify our understanding of granular media: Stress
relaxation is an elementary process, while plastic flows
are infamous for their complexity.

In a recent Letter~\cite{JL3}, some simplified equations
were derived based on the above guiding notion. For the
special case of a stationary $T_g$, these reproduce the
basic structure of hypoplasticity~\cite{Kolym-1}, a
state-of-the-art, rate-independent soil-mechanical model,
and yields an account of granular plastic flow that is
surprisingly realistic. As this agreement is a result of
fitting merely four numbers, we may with some confidence
take it as an indication that transient elasticity is
indeed a sound starting point, from which granular
hydrodynamics may be derived. It is not clear to us
whether this starting point alone is sufficient. More
work and exploration is needed, and especially cyclic
loading, critical state, shear banding and tapping need
to be considered. We reserve the study of these phenomena
for the future. In this paper, we take a first step in
our long march by deriving a consistent, hydrodynamic
framework (called {\sc gsh} for {\em granular solid
hydrodynamics}) starting from transient elasticity.

The paper is organized as follows. In
section~\ref{notion}, we discuss to what extent granular
media are elastic, or better, permanently elastic. It is
well known that, although the process leading to a given
granular state is typically predominantly plastic, the
excess stress field induced by a small external force in
a pre-stressed, static state can be described by the
equations of elasticity. We explain why, for $T_g=0$,
granular elasticity in fact extends well beyond this
limit, that it may be employed to calculate all static
stresses, not only incremental ones. The basic reason is,
without a finite $T_g$, there is no stress relaxation and
plastic flow. Similarly, if an incremental strain is
small enough, producing insufficient $T_g$, there is too
little plastic flow to mar the elasticity of a stress
increment.

Then we proceed, in section~\ref{GraEq}, to discuss {\em
jamming}, a word coined to describe a system prevented
from exploring the phase space, and confined to a single
state. Although this idea has proven rather
useful~\cite{jamming}, one must not forget that it is a
partial view, based on a truncated mesoscopic model, and
inappropriate for the present purpose. In this section,
jamming is generalized and embedded in the concept of
{\em constrained equilibria}. The point is, individual
grains are unlike atoms already macroscopic. They contain
innumerable internal degrees of freedom that are
neglected in mesoscopic
models~\cite{Haff-1,Haff-2,Haff-3,Haff-4,Haff-5}. For
instance, phonons contained in individual grains do
explore the phase space and arrive at a distribution
appropriate for the ambient temperature. Jamming fixes
only a few out of many, many degrees of freedom.
Realizing this, the fact that grains are prevented from
moving becomes comparable to the following textbook
example: Two chambers of different pressure, separated by
a {\em jammed} piston, and prevented from going to the
lowest-energy state of equalized pressure. Such a system
is in equilibrium and amenable to thermodynamics, albeit
under the constraint of two constant subvolumes.
Similarly, a jammed granular system at $T_g=0$ is also in
equilibrium, not in a single state, and amenable to
thermodynamics, although (as we shall see) under the
local constraint of a given density field
$\rho(\boldsymbol r)$ that cannot change even when
nonuniform. Exploring this analogy, section~\ref{GraEq}
arrives at a number of equilibrium conditions,  useful
both for describing granular statics and setting up
granular dynamics.

In section~\ref{Tg}, the physics of the granular
temperature is specified and developed. As mentioned, the
energy change ${\rm d}w$ from all microscopic, implicit
variables is usually subsumed as $T{\rm d}s$, with $s$
the entropy and $T\equiv\partial w/\partial s$ its
conjugate variable. From this, we divide out the
mesoscopic, intergranular degrees of freedom (such as the
kinetic and elastic energy of random, small-scaled
granular motion), denoting them summarily as the granular
entropy $s_g$. This is necessary, because these are
frequently rather more strongly agitated than the truly
microscopic ones, $T_g\equiv\partial w/\partial s_g\gg
T$. Note that in granular solids, we are equally
interested in the regime $T_g\gtrsim T$, as this is where
the elasticity switches from being transient to
permanent. In section~\ref{Tg}, the equilibrium condition
and equation of motion for $s_g$ are derived -- by taking
it to be an independent, macroscopic variable, without
any assumptions about how ``thermal" the associated
mesoscopic degrees of freedom are. (As mentioned, usually
they are not Gaussian and do not satisfy the
equipartition theorem.) However,  we do assume a two-step
irreversibility, that the energy only goes from the
macroscopic degrees of freedom to the mesoscopic,
intergranular ones summarized in $s_g$, and from there to
the microscopic, innergranular ones $s$. The final
subsection deals with a misconception that, because the
fluctuation-dissipation theorem ({\sc fdt}) in terms of
the granular temperature does not usually hold, neither
does the Onsager relation. The point is, the validity of
{\sc fdt} in terms of the true temperature is never in
question. And the Onsager relation only depends on the
latter.

In section~\ref{elaPla}, the equation of motion for the
elastic strain is elucidated, and shown to fully
determine the evolution of the plastic strain as well. In
section~\ref{GraFE}, an explicit expression for the free
energy $f$ is presented. This is necessary, because the
energy $w$, or equivalently the free energy $f$, are (as
discussed above) material quantities. As such, the free
energy must be found either by careful observation of
experimental data, an exercise in trial and error, or
more systematically, through simulation and microscopic
consideration. We proceed along the first line, making
use mainly of the jamming transition that occurs as a
function of $\rho,T_g,u_{ij}$, to find this expression.
Section~\ref{GHT1} presents the formal derivation of the
hydrodynamic theory. The resulting equations are then
applied to reproduce the hypoplastic model in
section~\ref{hypo}. Finally, section~\ref{conclu} gives a
brief summary.

\section{Sand -- a Transiently Elastic Medium
\label{notion}}

Granular media possess different phases that, depending
on the grain's ratio of elastic to kinetic energy, may
loosely be referred to as gaseous, liquid and solid.
Moving fast and being free most of the time, the grains
in the gaseous phase have much kinetic, but next to none
elastic,
energy~\cite{Haff-1,Haff-2,Haff-3,Haff-4,Haff-5}. In the
denser liquid phase, say in chute flows, there is less
kinetic energy, more durable deformation, and a rich
rheology that has been scrutinized
recently~\cite{chute-1,chute-2,chute-3}. In granular
statics, with the grains deformed but stationary, the
energy is all elastic. This state is legitimately
referred to as solid because static shear stresses are
sustained. If a granular solid is slowly sheared, the
predominant part of the energy remains elastic, and we
shall continue to refer to it as being solid.

When a granular solid is being compressed and sheared,
the deformation of individual grains leads to reversible
energy storage that sustains a static, elastic stress.
But they also jiggle and slide, heating up the system
irreversibly. Therefore, the macroscopic granular strain
field $\varepsilon_{ij}=u_{ij}+p_{ij}$ has two
contributions, an elastic one $u_{ij}$ for deforming the
grains, and a plastic one $p_{ij}$ for the rest. The
elastic energy $w_1(u_{ij})$ is a function of $u_{ij}$,
not $\varepsilon_{ij}$, and the elastic contribution to
the stress $\sigma_{ij}$ is given as
$\pi_{ij}(u_{ij})\equiv-
\partial w_1/\partial u_{ij}$. With the total and elastic
stress being equal in statics, $\sigma_{ij}=\pi_{ij}$,
stress balance $\nabla_j\sigma_{ij}=0$ may be closed with
$\pi_{ij}=\pi_{ij}(u_{ij})$, and uniquely determined
employing appropriate boundary conditions. Our
choice~\cite{J-L-1,J-L-2,J-L-3} for the elastic energy
$w_1=w_1(u_{ij})$ is
\begin{eqnarray}
w_1=\sqrt{\Delta }\left(\frac25 {\mathcal B} \Delta
^2+ {\mathcal A}u_s^2\right)\equiv{\mathcal
B}\sqrt{\Delta }\left(\frac25\Delta ^2+
\frac{u_s^2}\xi \right), \label{1-1}\\ \label{1-2}
\pi_{ij}\equiv-\frac{\partial w_1}{\partial u_{ij}}
=\sqrt\Delta({\cal B}\Delta\,\delta _{ij}-2{\cal A}\,
u_{ij}^0) +{\cal A} \frac{u_s^2}{2\sqrt\Delta}\delta
_{ij},
\end{eqnarray}
where $\Delta\equiv -u_{\ell\ell}$, $u_s^2\equiv
u^0_{ij}u^0_{ij}$, $u^0_{ij}\equiv u_{ij}-\frac13
u_{\ell\ell}\,\delta_{ij}$. Three classical cases: silos,
sand piles and granular sheets under a point load were
solved employing these equations, producing rather
satisfactory agreement with experiments~\cite{ge-1,ge-2}.
The elastic coefficient $\mathcal{B}$, a measure of
overall rigidity, is a function of the density $\rho$.
Assuming a uniform $\rho$ (hence a spatially constant
$\cal B$), the stress at the bottom of a sand pile is (as
one would expect) maximal at the center. But a stress dip
appears if an appropriate nonuniform density is assumed.
Because the difference in the two density fields are
plausibly caused by how sand is poured to form the piles,
this presents a natural resolution for the dip's history
dependence, long considered mystifying.

Moreover, the energy $w_1$ is convex only for
\begin{equation}\label{1-2a}
u_s/\Delta\le\sqrt{2\xi},\quad \text{or}\quad \pi
_s/P\le\sqrt{2/\xi},
\end{equation}
(where $P\equiv\frac13\pi _{\ell \ell }$, $\pi_s^2\equiv
\pi^0_{ij}\pi^0_{ij}$, $\pi^0_{ij}\equiv \pi_{ij}-\frac13
\pi_{\ell\ell}\,\delta_{ij}$,) implying no elastic
solution is stable outside this region. Identifying its
boundary with the friction angle of $28^\circ$
gives~\cite{ge-1,ge-2}
\begin{equation}\label{1-2b}
  \xi\approx5/3
\end{equation}
for sand. Because the plastic strain $p_{ij}$ is clearly
irrelevant for the static stress, one may justifiably
consider granular media at rest, say a sand pile, as
elastic.

If this sand pile is perturbed by periodic tapping at its
base, circumstances change qualitatively: Shear stresses
are no longer maintained, and the conic form degrades
until the surface becomes flat. This is because part of
the grains in the pile lose contact with one another
temporarily, during which their individual deformation
decreases, implying a diminishing elastic strain
$u_{ij}$, and correspondingly, smaller elastic energy
$w_1(u_{ij})$ and stress $\pi_{ij}(u_{ij})$. The system
is now elastic only for a transient period of time. The
typical example for {\em transient elasticity} is of
course polymer, and the reason for its elasticity being
transient is the appreciable time it takes to disentangle
polymer strands. Although the microscopic mechanisms are
different, tapped granular media display similar
macroscopic behavior, and share the same hydrodynamic
structure.

When being slowly sheared, or otherwise deformed,
granular media behaves similarly to being tapped, and
turn transiently elastic. This is because in addition to
moving with the large-scale shear velocity $v_i$, the
grains also slip and jiggle, in deviation of it. Again,
this allows temporary, partial unjamming, and leads to a
relaxing $u_{ij}$.

One does not have to assume that this deviatory motion
is completely random, satisfying equipartition and
resembling molecular motion in a gas. It suffices that
the elasticity turns transient the same way, no matter
what kind of deviatory motion is present. In either
cases, it is sensible to quantify this motion with a
scalar. Referring to it as the granular entropy or
temperature is suggestive and helpful. The granular
entropy $s_g$ thus introduced is an independent
variable of {\sc gsh}, with an equation of motion that
accounts for the generation of $T_g$ by shear flows,
and how the energy contained in $T_g$ leaks into heat.
Only when $T_g$ is large enough, of course, is
granular elasticity noticeably transient.

\section{Jamming and Granular Equilibria\label{GraEq}}
Liquid and solid equilibria are first described, then
shown to correspond to the unjammed and jammed equilibria
of granular media.

\subsection{Liquid Equilibrium}
In liquid, the conserved energy density
$w(s,\rho,g_i)$ depends on the densities of entropy
$s$, mass $\rho$, and momentum $g_i=\rho v_i$. The
dependence on $g_i$ is universal, given simply by
\begin{equation}\label{2-1}
w(s,\rho,g_i)=w_0(s,\rho)+g_i^2/(2\rho),
\end{equation}
leaving the rest-frame energy $w_0$ to contain the
material dependent part. Its infinitesimal change, ${\rm
d}w_0=({\partial w_0}/{\partial s}){\rm d}s+({\partial
w_0}/{\partial\rho}){\rm d}\rho$, is conventionally
written as
\begin{equation}\label{2-3}
{\rm d}w_0=T{\rm d}s+\mu{\rm d}\rho,
\end{equation}
by defining
\begin{equation}\label{2-2}
T\equiv{\partial w_0}/{\partial s}|_\rho,\quad
\mu\equiv{\partial w_0}/{\partial\rho}|_s.
\end{equation}
It is useful to note that given Eq~(\ref{2-1}), the
relation  ${\partial
w}/{\partial\rho}|_{s,g_i}\equiv\mu-v^2/2$ holds, hence
\begin{equation}\label{2-3a}
{\rm d}w=T{\rm d}s+(\mu-v^2/2){\rm d}\rho+v_i{\rm
d}g_i.
\end{equation}
Consider a closed system, of given volume $V=\int {\rm
d}^3r$, energy  $\int w{\rm d}^3r$, and mass $\int\rho\,
{\rm d}^3r$. Whatever the initial conditions, it will
eventually arrive at equilibrium, in which the entropy
$\int s{\rm d}^3r$ is maximal, or equivalently, at
minimal energy for given entropy, mass and volume. To
obtain the mathematical expression for this final state,
one varies $\int w{\rm d}^3r$ for given $\int s{\rm
d}^3r$ and $\int\rho\, {\rm d}^3r$, arriving at the
following {\em equilibrium conditions},
\begin{equation}\label{2-4}
\nabla_i T=0,\quad \nabla_i\mu=0.
\end{equation}
Being expressions for optimal distribution of entropy and
mass, these two conditions may respectively be referred
to as the thermal and chemical one.

In mathematics, Eqs~(\ref{2-4}) are referred to as the
Euler-Lagrange equations of the calculus of variation.
The calculation is given in Appendix~\ref{appA}. More
details may be found in~\cite{3cd}, in which three
additional conserved quantities: momentum $\int g_i{\rm
d}^3r$, angular momentum $\int (\boldsymbol
r\times\boldsymbol g)_i {\rm d}^3r$, and booster $\int
(\rho r_i-g_it){\rm d}^3r$ were also considered, adding a
motional condition,
\begin{equation}\label{2-4a}
v_{ij}\equiv(\nabla_iv_j+\nabla_jv_i)/2=0,
\end{equation}
and altering the chemical one to
$\partial_tv_i+\nabla_i(\mu-v^2/2)=0$. We focus on
Eqs~(\ref{2-4}) here.

Including gravitation, the energy is $\bar w_0=w_0+\phi$,
with $G_k=-\nabla_i\phi$ the gravitational constant
pointing downwards. The generalized chemical potential is
\begin{equation}\label{2-4c}
\bar\mu(\rho)\equiv\partial\bar
w_0/\partial\rho=\mu+\phi,
\end{equation}
while chemical equilibrium, $\nabla_i\bar\mu=0$, is
\begin{equation}
\nabla_i\mu=G_i.
\end{equation}
This implies a nonuniform density represents the optimal
mass distribution minimizing the energy (or maximizing
the entropy). With the pressure given as
$P_T=-w_0+TS+\mu\rho$, see  Appendix~\ref{appA}, the
condition for mechanical equilibrium,
\begin{equation}
  \nabla_iP_T=s\nabla_iT+\rho\nabla_i\mu=\rho G_i
\end{equation}
is a combination of the thermal and chemical ones.

\subsection{Solid Equilibrium}
In solids, if the subtle effect of mass defects is
neglected, density is not an independent variable and
varies with the strain (for small strains) as
\begin{equation}\label{2-6}
{\rm d}\rho/\rho=-{\rm d}u_{\ell\ell}.
\end{equation}
Defining $\pi_{ij}\equiv-\partial w_0/\partial
u_{ij}|_s$, we write the change of the energy as
\begin{equation}\label{2-7}
{\rm d}w_0(s,u_{ij})=T{\rm d}s-\pi_{ij}{\rm d} u_{ij}.
\end{equation}
Maximal entropy, with the displacement vanishing at the
system's surface, implies the following thermal and
mechanical equilibrium conditions (see
Appendix~\ref{appA}),
\begin{equation}\label{2-8}
\nabla_iT=0,\quad \nabla_j\pi_{ij}=0.
\end{equation}
So force balance is, in the complete world including the
innergranular degrees of freedom, an expression of
maximal entropy -- quite analogous to uniformity of
temperature. It implies the overwhelming dominance of
phonon distribution that satisfies force balance, and the
rarity of phonon fluctuations that violate it.

Including gravitation, the total energy is given as ${\rm
d}\bar w_0(s,u_{ij})=T{\rm d}s-\bar\pi_{ij}{\rm d}
u_{ij}$, with $\bar\pi_{ij}=\pi_{ij}+\rho\phi$, and
mechanical equilibrium becomes
\begin{equation}
\nabla_j\pi_{ij}=\rho G_i
\end{equation}

\subsection{Granular Equilibria\label{granular equilibria}}
Depending on whether $T_g$ is zero or finite, sand
flip-flops between the above two types of behavior. The
density is an independent variable, because the grains
may be differently packaged, leading to a density
variation of between 10 and 20\% at vanishing
deformation. So the energy depends on all three
variables,
\begin{equation}\label{2-10}
{\rm d}w_0(s,\rho,u_{ij})=T{\rm d}s+\mu{\rm
d}\rho-\pi_{ij}{\rm d} u_{ij}.
\end{equation}
If $T_g$ is finite, the elastic stress $\pi_{ij}$ relaxes
until it vanishes. The equilibrium conditions are
therefore, including gravitation,
\begin{equation}\label{2-12}
\nabla_i T=0,\quad \nabla_i P_T=\rho\, G_i,\quad
\pi_{ij}=0,
\end{equation}
similar to that of a liquid, with $\nabla_i P_T=\rho\,
G_i$ (or $\nabla_i\mu=G_i$) enforcing an appropriate
density field, and $\pi_{ij}=0$ forbidding any free
surface other than horizontal.

For vanishing $T_g$, sand is jammed, implying two points:
First, $\pi_{ij}$ no longer relaxes; second, without
slipping and jiggling, the packaging density cannot
change, and the density is again a dependent variable,
${\rm d}\rho/\rho=-{\rm d}u_{\ell\ell}$. The suitable
equilibrium conditions, as derived in
Appendix~\ref{appA}, are
\begin{equation}\label{2-11}
\nabla_iT=0,\quad \nabla_j(P_T\delta_{ij}+\pi_{ij})=\rho
G_i,
\end{equation}
which allow static shear stresses and tilted free
surfaces. So, although jammed states are prevented from
arriving at the liquid-like conditions of
Eqs~(\ref{2-12}), they do possess reachable thermal and
mechanical equilibria.

If the energy (as given in section~\ref{Tg}) depends in
addition on the granular entropy, ${\rm d}w=T{\rm
d}s+T_g{\rm d}s_g+\cdots$, the pressure contribution
$P_T$ (see section~\ref{GHT}) is
\begin{eqnarray}\label{app32a}
P_T=-w_0+Ts+T_gs_g+\mu\rho=-\tilde f+\mu\rho,
\\
\text{with}\quad\nabla_iP_T=s\nabla_iT+s_g\nabla_iT_g+\rho\nabla_i\mu.
\end{eqnarray}

\section{Granular Temperature $\boldsymbol {T_g}$
\label{Tg}}

Granular temperature is not a new concept. Haff, at the
same time Jenkins and
Savage~\cite{Haff-1,Haff-2,Haff-3,Haff-4,Haff-5},
introduced it in the context of granular gas, taking (in
an analogy to ideal gas) $T_g\sim w_{\rm kin}$, where
$w_{\rm kin}$ is the kinetic energy density of the grains
in a quiescent granular gas. With $T_g\equiv\partial
w_{\rm kin}/\partial s_g\sim\partial T_g/\partial s_g$,
the granular entropy is $s_g\sim \ln T_g$. As discussed
above, granular temperature is also a crucial variable in
granular solids. But one must not expect this gas-like
behavior to extend to the vicinity of  $T_g=T$: As the
system, if left alone, always returns to  $T_g=T$, the
energy must have a minimum there. And something like
$w\sim s_g^2\sim(T_g-T)^2$ and $s_g\sim T_g-T$ would be
more appropriate. (Neither for ideal gases does $s_g\sim
\ln T_g$ persist for all temperature. Excluding a phase
transition, quantum effects become important before $T=0$
is reached.)
\subsection{The Equilibrium Condition for $\boldsymbol{T_g}$}
The energy change ${\rm d}w$ from all microscopic,
implicit variables is generally subsumed as $T{\rm d}s$,
with $s$ the entropy and $T\equiv\partial w_0/\partial s$
its conjugate variable. From this, we divide out the
intergranular energy of the random motion of the grains,
denoting it as $T_g{\rm d}s_g$,
\begin{equation}\label{2-13}
{\rm d}w_0=T{\rm d}(s-s_g)+T_g{\rm d}s_g=T{\rm
d}s+(T_g-T){\rm d}s_g.
\end{equation}
The first expression distinguishes between two heat
pools: $s-s_g$ and $s_g$, with the latter rather more
strongly excited, $T_g\gg T$. The second expression,
algebraically identical, takes $w$ as a function of $s$
and $s_g$, with $T{\rm d}s$ being the total heat if all
degrees were at $T$, and $(T_g-T){\rm d}s_g$ the increase
in energy when some of the degrees are at $T_g$. If
unperturbed, a stable system will always return to
equilibrium, at which the second pool is empty, $s_g=0$.
This implies the free energy $f\equiv w_0-Ts$ has a
minimum at $s_g=0$. Assuming analyticity, we expand the
free energy $f(T,s_g)$ around $s_g=0$, arriving at
\begin{equation}f=f_0(T)+s_g^2/(2b\rho),\label{2-14d}
\end{equation}
where $b$ is a positive material parameter, a function of
$\rho$ and $u_{ij}$. [The factor $\rho$ will turn out
later to be convenient.] With ${\rm d}f=-s{\rm
d}T+(T_g-T){\rm d}s_g$ we have
\begin{equation}
\label{2-15} \bar T_g\equiv T_g-T\equiv \partial
f/\partial s_g|_{T} =s_g/(b\rho)
\end{equation}
that vanishes in equilibrium
\begin{eqnarray}
\bar T_g\equiv T_g-T=0. \label{2-22}
\end{eqnarray}
We shall employ the Legendre transformed potential,
$\tilde f(T,\bar T_g)\equiv f(T,s_g)-\bar T_gs_g$, below
(that has a maximum rather than a minimum at $T_g=T$),
\begin{equation}\tilde f(T,\bar T_g)=
f_0(T)-b\rho\bar T_g^2/2.\label{2-14}
\end{equation}

Because an improbably high $T_g$ is implied by any random
motion of the grains, neglecting $T$ in comparison to
$T_g$ or taking $\bar T_g\approx T_g$ is frequently a
good approximation, though not close to $\bar T_g=0$. So
it is prudent not to implement it while deriving the
equations.

\subsection{The Equation of Motion for $\boldsymbol{s_g}$}
Being a macroscopic, non-hydrodynamic variable, $s_g$
must first of all obey a relaxation equation,
$-{\partial_t}s_g =\gamma\partial f/\partial s_g=\gamma
\bar T_g$. Since this relaxation is typically slow, $s_g$
also displays characteristics of a quasi-conserved
quantity, and removal of local accumulations is accounted
for by a convective and a diffusive term,
\begin{eqnarray}\label{2-16}
-{\partial_t}s_g&=&\nabla_i[s_gv_i-\kappa_g\nabla_i\bar
T_g]+\gamma \bar
T_g\\&=&\nabla_i(s_gv_i)+(1-\chi^2\nabla^2)s_g/\tau_g,
\nonumber\end{eqnarray}
where $\tau_g\equiv b\rho/\gamma$ is the relaxation time,
while $\chi\equiv\sqrt{\kappa_g/\gamma}$ is the
characteristic length associated with the diffusion. (The
second line of Eq~(\ref{2-16}) assumes $\kappa_g,\gamma=$
constant.) If $\bar T_g$ is held at $T_0$ at the boundary
$x=0$, and allowed to relax for $x>0$, the field
$s_g\sim\bar T_g(x)$ obeys $(1-\chi^2\nabla^2)s_g=0$ in
the stationary limit ${\partial_t}s_g, v_i=0$, and decays
as
\begin{equation}\label{2-22a}
T_g(x)=T+T_0\exp(-x/\chi).
\end{equation}

Eq~(\ref{2-16}) is not complete. To see this, consider
first the true entropy $s$. In liquid, $s$ is governed by
a balance equation with a positive source term $R$ that
is fed by shear and compressional flows, and by
temperature gradients~\cite{LL6},
\begin{eqnarray}\label{2-17}
\partial_ts+\nabla_i(sv_i-\kappa\nabla_iT)
=R/T,
\\R=\eta v_{ij}^0v_{ij}^0+\zeta v_{\ell\ell}^2+
\kappa(\nabla_iT)^2, \label{2-18}
\end{eqnarray}
where $v_{ij}^0$ is the traceless part of
$v_{ij}\equiv\frac12(\nabla_iv_j+\nabla_jv_i)$ and
$v_{\ell\ell}$ its trace; $\eta,\zeta>0$ are the shear
and compressional viscosity, respectively, and $\kappa>0$
the heat diffusion coefficient. Entropy production  $R$
must vanish in equilibrium and be positive definite off
it. The {thermodynamic forces} $\nabla_iT$ and $v_{ij}$
also vanish in equilibrium [see
Eqs~(\ref{2-4},\ref{2-4a})]; off it, they may be taken to
quantify the ``distance from equilibrium." The entropy
production $R$ increases with this distance and may be
expanded in  $\nabla_iT$ and $v_{ij}$. The given terms
are the lowest order, positive ones that are compatible
with isotropy.

In granular media, equilibrium conditions are more
numerous than in liquid. As discussed in
section~\ref{granular equilibria}, these are, in
addition, the vanishing of $\pi_{ij}$,
$\nabla_j\pi_{ij}$, and $\bar T_g$, hence we have
\begin{eqnarray}\label{2-18a}
R=\eta v_{ij}^0v_{ij}^0+\zeta v_{\ell\ell}^2+
\kappa(\nabla_iT)^2+\gamma \bar T_g^2
\\\nonumber
+\beta(\pi^0_{ij})^2+\beta_1\pi_{\ell\ell}^2
+\beta^P(\nabla_j\pi_{ij})^2.
\end{eqnarray}
Three additional points: (1)~Being an expansion in the
thermodynamic forces, the transport coefficients $\eta,
\zeta, \kappa, \kappa_g,  \gamma, \beta, \beta_1,
\beta^P$ may still depend on the variables of the energy,
$T, \bar T_g, \rho$, $\pi_{\ell\ell}$ and $\pi_s^2\equiv
\pi_{ij}^0\pi_{ij}^0$, but not on the forces themselves,
such as $\nabla_iT$ or $v_{ij}$.  (2)~More terms are
conceivable in Eq~(\ref{2-18a}), say
$\alpha_1\nabla_iT\nabla_j\pi_{ij}$ or
$\kappa_1\pi_{ij}\nabla_iT\nabla_jT$. These may be
included when necessary.  (3)~The above reasoning leaves
the question open why $\nabla_i\mu$ does not contribute
to $R$, not even in liquid -- or more precisely, why the
coefficient preceding $(\nabla_i\mu)^2$ always vanishes.
The answer is given in~\cite{3cd}, though there have been
some recent controversies about it, see~\cite{oett} and
references therein.

The granular entropy $s_g$ should obey a balance equation
with the same structure,
\begin{equation}\label{2-19}
\partial_ts_g+\nabla_i(s_gv_i-\kappa_g\nabla_i\bar T_g)
=R_g/\bar T_g,
\end{equation}
though the source term $R_g$ has positive as well as
negative contributions: Two positive ones from shear and
compressional flows, and the negative relaxation term
discussed in Eq~(\ref{2-16}),
\begin{equation}R_g=\eta_g v_{ij}^0v_{ij}^0
+\zeta_g v_{\ell\ell}^2+ \kappa_g(\nabla_i\bar
T_g)^2-\gamma \bar T_g^2. \label{2-20}
\end{equation}
The fact that the coefficient preceding $\bar T_g^2$ is
$\gamma$ both in Eq~(\ref{2-18a}) and (\ref{2-20})
derives from energy conservation: Taking the system to be
uniform, we have $\partial_tw=T\partial_ts+\bar
T_g\partial_ts_g =R+\bar T_g(-\gamma\bar T_g)$. So
$\partial_tw=0$ implies $R=\gamma\bar T_g^2$. It
expresses the fact that the same amount of heat leaving
$s_g$ must arrive at $s$. A direct consequence for the
stationary case, $R_g=0$, is
\begin{equation}\label{2-21}
\gamma\bar T_g^2=\eta_g v_{ij}^0v_{ij}^0+\zeta_g
v_{\ell\ell}^2,
\end{equation}
quantifying how much $\bar T_g\equiv T_g-T$ is excited by
shear or compressional flows.

In dry sand, the granular viscosities $\eta_g,\zeta_g$
probably dominate, while $\eta,\zeta$ are insignificant
-- though the latter should be quite a bit larger in sand
saturated with water: A macroscopic shear flow of water
implies much stronger microscopic ones in the fluid
layers between the grains, and the energy dissipated
there goes to the true entropy $s$, instead of to $s_g$
first.

\subsection{Two Fluctuation-Dissipation Theorems\label{onsager}}

There are many in the granular community who dispute the
validity of the Onsager reciprocity relation in granular
media, enlisting any of the following three reasons:
(1)~The fluctuation-dissipation theorem ({\sc fdt}) does
not hold. (2)~The microscopic dynamics is not reversible.
(3)~Sand is too far off equilibrium.

Careful scrutiny shows that none of these arguments holds
water. First, with $F$ denoting the free energy,
fluctuations say of the volume are always given as
\begin{equation}\label{GSH-k1}
 \langle\Delta V^2\rangle=T(\partial^2 F/\partial
V^2)^{-1}=T(-\partial P/\partial V)^{-1}.
\end{equation}
Jammed sand, similar to a copper block, undergoes volume
fluctuations as described by Eq~(\ref{GSH-k1}). When sand
is unjammed, Eq~(\ref{GSH-k1}) still holds, though $F$
now depends on $T_g$, such as given in
section~\ref{GraFE}. In granular media, $T$ is frequently
replaced by $T_g$,
\begin{equation}\label{GSH-k2}
\langle\Delta V^2\rangle=T_g(-\partial P/\partial
V)^{-1}.\end{equation}
This ``{\sc fdt}" is indeed highly questionable, because
$T_g$ frequently behaves rather differently from the true
temperature. However, the crucial point here is, the
validity of the Onsager relation depends on
Eq~(\ref{GSH-k1}), not Eq~(\ref{GSH-k2}).

Second, the dynamics typically employed in granular
simulations is indeed irreversible, but only as a result
of a model-dependent approximation that treats grains as
elementary constituent entities. The true microscopic
dynamics that resolves the atomic building blocks of the
grains remains reversible. And this is the basis for the
Onsager relation.

Third, ``too far off equilibrium" is not convincing, as
turbulent fluids, truly far off equilibrium, are known to
obey them. Some argue that sand, whether  jammed or in
motion,  are always far from equilibrium. Yet as the
careful discussion in section~\ref{GraEq} shows, this is
an inappropriate view. Granular media are not always far
from equilibrium, they just have different ones to go to
-- solid-like if jammed and liquid-like if unjammed.

\section{Elastic and Plastic Strain\label{elaPla}}

As discussed in section~\ref{notion}, the elastic
strain $u_{ij}$ accounts for the deformation of
individual grains, while their rolling and sliding is
described by the plastic strain $p_{ij}$. Together,
they form the total strain
\begin{equation}
\varepsilon_{ij}= u_{ij}+p_{ij}.
\end{equation}
The elastic energy $w(u_{ij})$ is a function of
$u_{ij}$, not of $\varepsilon_{ij}$, and the elastic
stress is given as $\pi_{ij}(u_{ij})\equiv-
\partial w/\partial u_{ij}$. When $T_g$ is finite, the
elastic strain relaxes,
\begin{equation}\label{xx2-22}
\partial_t u_{ij}-v_{ij}=-u_{ij}/\tau.
\end{equation}
implying a diminishing elastic strain $u_{ij}$, and
correspondingly, smaller elastic energy $w(u_{ij})$ and
stress $\pi_{ij}(u_{ij})$. Note because the total strain
is a purely kinematic quantity,
$\partial_t\,\varepsilon_{ij}=v_{ij}$, the evolution of
the plastic strain $p_{ij}$ is also fixed, $\partial_t\,
p_{ij}= v_{ij}-\partial_t u_{ij}$.

It is the relaxation term $-u_{ij}/\tau$ that gives rise
to plasticity. To see how it works, take a constant
$\tau$ and consider the following scenario. If a
transiently elastic medium is deformed quickly enough by
an external force, leaving little time for relaxation,
$\int (u_{ij}/\tau)\,{\rm d}t\approx0$, we have
$\varepsilon_{ij}=u_{ij}$ and $p_{ij}=0$ right after the
deformation. If released at this point, the system would
snap back toward its initial state, as prescribed by
momentum conservation, $\partial_t\,(\rho
v_i)+\nabla_j\pi_{ij}=0$, displaying thus a behavior that
is clearly reversible and elastic. But if we hold the
system still for long enough, $v_{ij}=0$, hence
$\partial_t\,\varepsilon_{ij}=0$, the elastic part
$u_{ij}$ will relax, $\partial_tu_{ij}=-u_{ij}/\tau$,
while the plastic part grows accordingly,
$\partial_tp_{ij}= -\partial_tu_{ij}$. When $u_{ij}$
vanishes, the plastic part will have completely replaced
it, $p_{ij}=\varepsilon_{ij}$. With the elastic stress
$\pi_{ij}$ and energy $w(u_{ij})$ also gone, momentum
conservation reads $\partial_t\,(\rho v_i)=0$. The system
now stays where it is when released, and no longer strive
to return to its original position. This is obviously
what we mean by a plastic deformation.

Next take $\tau\sim T_g^{-1}$. As discussed in the
introduction, this should be appropriate for granular
media. Assuming (for simplicity) a stationary granular
temperature, or
$T_g^2=(\eta_g/\gamma){v_{ij}v_{ij}}\equiv(\eta_g/\gamma)||v_{s}||^2$,
see Eq~(\ref{2-21}), we obtain from Eq~(\ref{xx2-22})
the equation,
\begin{equation}\label{xx2-222}
\partial_t
u_{ij}-v_{ij}\sim||v_{s}||(-u_{ij})\sqrt{\eta_g/\gamma}\,,
\end{equation}
the rate-independent structure of which closely resembles
the hypoplastic one~\cite{Kolym-1}. As a result, both the
elastic strain $u_{ij}$ and the stress $\sigma_{ij}$ will
display {\em incremental nonlinearity}, ie., behave
differently depending whether the load is being increased
($v_{ij}>0,\,\,||v_{s}||>0$) or decreased
($v_{ij}<0,\,\,||v_{s}||>0$). Not surprisingly, this
equation leads to plastic flows very similar to the
hypoplastic results. However, under cyclic loading of
small amplitudes, because $T_g$ never has time to grow to
its stationary value, the plastic term $u_{ij}/\tau\sim
T_gu_{ij}$ remains small, and the system's behavior is
rather more elastic.

The equation of motion for the elastic strain [cf. the
derivation leading to Eq~(\ref{15y-GSH})] is in fact
somewhat more complicated and given as
\begin{eqnarray}\nonumber {\rm d}_t
u_{ij}-(1-\alpha)v_{ij}-X_{ij}\qquad\qquad\qquad
\\=-[(u_{ik}\nabla_{j}v_k+\nabla_iy_j/2)
+(i\leftrightarrow j)], \label{2-25}
\end{eqnarray}
where ${\rm d}_t\equiv\partial_t+v_k\nabla_k$, and
$(i\leftrightarrow j)$ signifies the same expressions as
in the preceding bracket, only with the indices $i$ and
$j$ exchanged. In this equation, the term
$(u_{ik}\nabla_{j}v_k)+(i\leftrightarrow j)$, important
for large strain field and frequently negligible for hard
grains, is of geometric origin,
see~\cite{polymer-1,polymer-2,polymer-3,polymer-4} for
explanations. The  dissipative fluxes
$X_{ij}=-u_{ij}/\tau-\alpha v_{ij}$ and $y_i\sim
\nabla_j\pi_{ij}$ will be derived in section~\ref{GHT}.
The second term is quite similar to the diffusive heat
current $\kappa\nabla_iT$, which aims to reduce
temperature gradients and establish $\nabla_iT=0$. We can
take $y_i$ to be a current that aims to reduce
$\nabla_j\pi_{ij}$ and establish the equilibrium
condition, $\nabla_j\pi_{ij}=0$, of Eq~(\ref{2-11}).
Given Eq~(\ref{2-25}) and ${\rm
d}_t\,\varepsilon_{ij}+[(\varepsilon_{ik}\nabla_{j}v_k)
+(i\leftrightarrow j)]-v_{ij}=0$, the evolution for the
plastic strain is again fixed.

\section{The Granular Free Energy\label{GraFE}}

As explained in the Introduction, the structure of the
hydrodynamic theory is determined by general principles,
especially energy and momentum conservation, but the
explicit form of the energy $w$ is not. Although $w$ does
possess features that it must always satisfy, most of its
functional dependence reflects the specific behavior of
the material. To arrive at an expression for the energy
of granular media, there are two obvious methods, either
a microscopic derivation, possibly via simulation, or
more pragmatically, examining constraints from key
experiments, opting for simplicity whenever possible, as
we do here.

Because we are interested in the limit of small $T_g$ and
$u_{ij}$, see Eq~(\ref{1-1}) and (\ref{2-14d}), and
because the dependence on the true temperature is usually
irrelevant, the difficult part is the density dependence
of the energy. Fortunately, quite a number of known
features may be used as input. First, there are two
characteristic granular densities, the minimal and
maximal ones, $\rho_{\ell p}$ and $\rho_{cp}$,
respectively referred to as {\it random loosest} and {\it
closest packing}. In the first case, the grains
necessarily loose contact with one another when the
density is further decreased; in the second, the density
can no longer be increased without compression, at which
point the system is orders of magnitude
stiffer~\cite{elaPla-1,elaPla-2,elaPla-3,Liniger}. Then
there is the jamming transition of sand, especially the
so-called {\em virgin consolidation line}, which we
believe is the limit beyond which no stable elastic
solutions are possible, see Fig~\ref{fig1}-(a). These in
conjunction with the density dependence of sound velocity
and the pressure exerted by agitated grains contain
sufficient information to fix the expression for the
energy.

Instead of the energy, we consider the potential $\tilde
f(T, \bar T_g, \rho, u_{ij})\equiv w_0-Ts-\bar T_gs_g$,
see Eq~(\ref{2-14}). Referring to it for simplicity also
as the free energy density, we write
\begin{eqnarray}\label{fe1}
\tilde f&=&f_0(T,\rho)+f_1(\rho,u_{ij})+f_2(\rho,\bar
T_g),
\\\label{fe3}
f_1&\equiv&w_1={\cal B}\sqrt{\Delta}
\,\,(2\Delta^2/5+u_s^2/\xi),
\\
\label{fe2}f_2&=&\rho\,b_0(1-\rho/\rho_{cp})^{a}(-\bar
T_g^2/2),\quad 0<a\ll1,
\end{eqnarray}
where  $f_0(T,\rho)$ is the free energy at vanishing
granular temperature and elastic deformation, $\bar T_g,
u_{ij}=0$, while $w_1(u_{ij})$ and $f_2(\bar T_g)$ are
the respective lowest order term. (It is a simplifying
assumption that the temperature $T$ enters the free
energy only via $f_0$, and not $w_1,f_2$. This neglects
effects such as thermal expansion that, however, may be
added when necessary.)

Being cohesionless, the grains possess no interaction
energy, $f_0(T,\rho)$ is therefore the sum of the free
energy in each of the grains,
\begin{equation}\label{fe3a}
f_0(T,\rho)=\langle F_1(T)/m\rangle\rho,
\end{equation}
where $F_1$ is the free energy of a single grain, $m$ its
mass, and $\langle F_1(T)/m\rangle$ the free energy per
unit mass, averaged over a number of grains.

It is important to realize that the equilibrium stress is
given, once one knows what the free energy density
$\tilde f=F/V$ is (see Appendix~\ref{appA}),
\begin{equation}\label{fe4b}
\sigma_{ij}=P_T\delta_{ij}+\pi_{ij}=-\left[\frac{\partial
(\tilde f/\rho)}{\partial(1/\rho)}\right]\delta_{ij}
-\frac{\partial\tilde f}{\partial u_{ij}}.
\end{equation}
The first term is the local expression for the more
familiar one,
\begin{eqnarray}\label{fe4a}\nonumber
P_T\equiv-\frac{\partial F}{\partial
V}=-\left.\frac{\partial (\tilde
fV/M)}{\partial(V/M)}\right|_M=-\frac{\partial (\tilde
f/\rho)}{\partial(1/\rho)}\\=\rho\partial\tilde
f/\partial\rho-\tilde f=\rho\mu +Ts+\bar T_gs_g-w.
\end{eqnarray}
In liquids, only this term exists, since $\tilde f$ does
not depend on $u_{ij}$; in ideal crystals,  only the
second term exists, because the density is not an
independent variable, see the discussion in section
\ref{GraEq}. In granular media, both terms coexist. Given
the free energy $\tilde f=\sum f_i$ of Eq~(\ref{fe1}),
each term yields the pressure contribution,
\begin{equation}\label{fe4}
P_i\equiv\rho(\partial f_i/\partial\rho)-f_i,
\end{equation}
with $P_T\equiv\sum P_i$ and $P_0\equiv\rho\partial
f_0/\partial\rho-f_0=0$.

\subsection{The Elastic Energy\label{Elastic}}

The elastic part of the free energy, Eq~(\ref{fe3}), has
previously been successfully tested under varying
circumstances, cf. the discussion in
section~\ref{notion}, below Eq~(\ref{1-2}). It is not
analytic in the elastic strain, but does contain the
lowest order terms. As it takes some deliberation to
arrive at its density dependence and the terms of higher
order in $u_{ij}$, we consider them in two separate
sections below.
\begin{figure}[b]
\begin{center}
\includegraphics[scale=0.8]{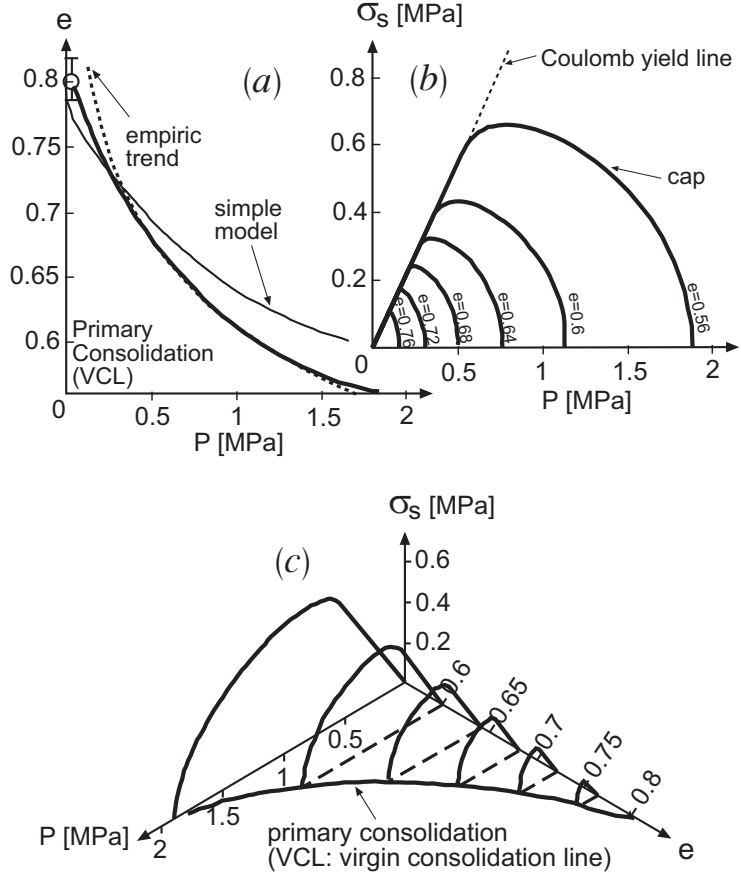}
\end{center}
\caption{\label{fig1}Granular yield surface, or
jamming phase diagram, for $T_g=0$, as a function of
the pressure $P$, shear stress $\sigma_s$, and void
ratio $e\equiv\rho_G/\rho-1$. All thick solid lines
are calculated using
Eqs~(\ref{fe3},\ref{fe10},\ref{fe12a}). (a):~Maximal
void ratio $e$ versus pressure $P$, or the {\em virgin
consolidation line.} The dotted line is an empirical
formula, $e=0.679-0.097\ln (P/0.5)$, with $P$ in Mpa.
The thin line (designated as {\em simple model})
renders Eq~(\ref{fe11}). The circle at the top is the
random loosest packing value for $e$. (b):~The
straight Coulomb yield line bends over depending on
$e$, a behavior usually accounted for by the {\em cap
model} in elasto-plastic theories. (c):~The 3D
combination of (a) and (b). Values for the calculation
are :~${\cal B}_0=7000$ Mpa, $\rho _{\ell
p}^*=0.445\rho _G$, $\rho _{cp}=0.645\rho _G$, $\Delta
_1=10^{-4}$, and $k_1=10^{-5}$ m$^3$/kg, $k_2=1000$,
$k_3=0.01$. }
\end{figure}

First, a conceptual point. We take any yield surface as
the divide between two regions: One in which stable
elastic solutions are possible, the other in which they
are not -- so a system under stress must flow and cannot
come to rest here. Accepting this, the natural approach
is to have a convex elastic energy turn concave at the
yield surface. The idea behind it is, the energy is an
extremum if the equilibrium conditions of
section~\ref{GraEq}, including especially
Eq~(\ref{2-11}), are met. Convexity implies the energy is
at a minimum there, and concavity that it is at a
maximum. Where $w_1$ is concave, any elastic solution
satisfying Eq~(\ref{2-11}) has maximal energy, and is
eager to get rid of it. It is not stable because
infinitesimal perturbations suffice to destroy it.

As discussed in section~\ref{notion}, for ${\cal B},\xi$=
constant, $w_1$ is convex for $\pi _s/P\le\sqrt{2/\xi}$
and concave otherwise, and already possesses the right
form to account for the Coulomb yield line, see
Fig~\ref{fig1}-(b). Our task now is to appropriately
generalize it such that the density $\rho$ is included as
a third variable. Instead of $\rho$, the void ratio,
$e\equiv\rho_G/\rho-1$, is frequently employed. It
remains constant at elastic compressions and accounts for
granular packaging only. ($\rho_G$ the bulk density of
granular material, typically around $2700$ kg/m$^3$ for
sand.)

\subsection{Density Dependence of $\cal B$\label{loose}}
We shall take ${\cal B}$ as density dependent, but not
$\xi$: Since the Coulomb yield line is approximately
independent of the density, so must the coefficient $\xi$
be, see Eq~(\ref{1-2b}). Granular sound velocity was
measured by Hardin and Richart~\cite{Hardin}, who found
it linear in the void ratio, $c\sim 2.17-e$. Given
Eq~(\ref{fe3}), the velocity of sound is
$c\sim\sqrt{{\cal B}/\rho}$, implying
\begin{equation}\label{fe8}
{\cal B}={\cal B}_0 (3.17-\rho_G/\rho)^2(\rho/\rho_G).
\end{equation}

Since this expression properly accounts for the
measured~\cite{Kuwano} density dependence of the
compliance tensor $M_{ijk\ell}$, the dependence of
$\cal B$ on $\rho$ seems settled~\cite{SoilMech}. It
is not, because the resultant $w_1$ is concave in the
variables $\rho$ and $\Delta$, and could not possibly
sustain any static solution. Inserting Eq~(\ref{fe8})
into (\ref{fe3}), we find the energy violating the
stability condition,
\begin{equation}\label{fe9}
 \partial^2\mathcal{B}^{-2/3}/\partial \rho ^2\leq 0,
\end{equation}
obtained from inserting Eq~(\ref{fe3}) with
$u_s\equiv0$ into $(\partial ^2w_1/\partial \rho ^2)\,
(\partial ^2w_1/\partial \Delta ^2) \geq (\partial
^2w_1/\partial \rho
\partial \Delta )^2$.
Clearly, the widely employed Hardin-Richart relation,
$c\sim 2.17-e$, is not accurate enough for a direct input
into the energy. It works fine as long as the sand is
jammed, $T_g=0$, and $\rho$ is only a given parameter,
not a free variable -- such as in the experiments
of~\cite{Kuwano}, or when determining static stress
distributions. But if a finite $T_g$ frees the density to
become a variable, this instability will wreck havoc with
the hydrodynamic theory. We need to reconstruct the
density dependence of $\cal B$, such that the energy
$w_1$
\begin{enumerate}
\item vanishes for densities smaller than
the random loosest packing value (around the void ratio
of $e_{\ell p}\approx0.8$ for sand of uniform grain
size), or $\rho\leq\rho_{\ell p}$;
\item (as a simplification) diverges
at $\rho=\rho_{cp}$, the random closest packing value
(around $e_{cp}\approx0.55$);
\item is convex and reproduces the
Hardin-Richart relation between $\rho_{\ell p}$ and
$\rho_{cp}$.
\end{enumerate}
\begin{figure}[b]
\begin{center}
\includegraphics[scale=0.6]{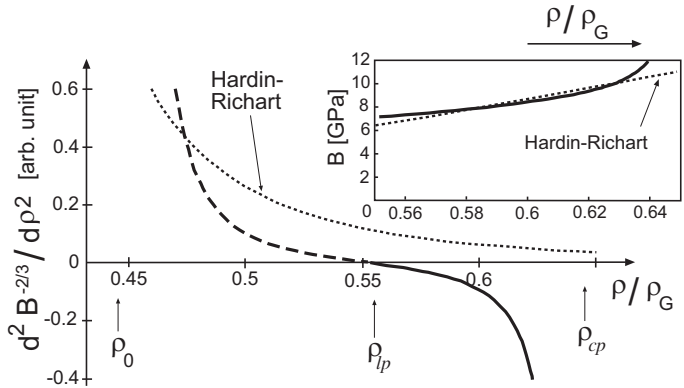}
\end{center}
\caption{Equation~(\ref{fe8}), obtained by employing the
Hardin-Richart relation directly, violates the stability
condition Eq~(\ref{fe9}), because
$\partial^2\mathcal{B}^{-2/3}/\partial \rho ^2>0$ for all
density values. Although numerically similar, see insert,
the expression from Eq~(\ref{fe10}) suitably becomes
concave at $\rho _{\ell c}$, and satisfies the stability
condition between $\rho _{\ell c}$ and $\rho_{cp}$. The
plots are calculated with $\rho _{\ell c}^*=0.445
\rho_G$, $\rho _{pc}=0.645\rho_G$ (implying $\rho_{\ell
p}=0.555\rho_G$), and ${\cal B}_0=7000$ Mpa, appropriate
for Ham River sand~\cite{Kuwano}. } \label{HR}
\end{figure}
Alas, these points are more easily stated than
combined in an energy expression, and no continuous
$\cal B$ seems feasible: If analytic, $\cal B$ would
be proportional to $\rho-\rho_{\ell p}$ close to
$\rho_{\ell p}$. More generally, we may take ${\cal
B}\sim(\rho-\rho_{\ell p})^\alpha$, with $\alpha$
positive. But the resulting energy,
$w\sim(\rho-\rho_{\ell p})^\alpha\Delta^{2.5}$,
remains concave. Only when including the divergence at
$\rho_{cp}$ by taking ${\cal B}\sim(\rho-\rho_{\ell
p})^\alpha /(\rho_{cp}-\rho)^\beta$ does the energy
turn convex, between $\rho_{cp}$ and a density larger
than $\rho_{\ell p}$. We therefore propose
\begin{eqnarray}\label{fe10} {\cal
B}&=&{\cal B}_0\left(\frac{\rho -\rho^*_{\ell p}}{\rho
_{cp}-\rho }\right) ^{0.15}\times{\cal C},\quad
\text{for}\,\, \rho>\rho_{\ell p};
\\
{\cal B}&=&0,\qquad\qquad\qquad\qquad\quad
 \text{for}\,\,\rho\leq\rho_{\ell p}.
\end{eqnarray}
With an appropriate $\rho^*_{\ell p}<\rho_{\ell p}$, this
expression renders the energy divergent at $\rho_{cp}$,
stable and convex up to $\rho_{\ell p}$, and approximates
the Hardin-Richart relation between them, see
Fig.~(\ref{HR}). (Take ${\cal C}=1$ for now, until it is
specified otherwise in the next section.)


\subsection{Higher-Order Strain Terms\label{virgin}}

Next, we consider the unjamming transition in connection
with {\it compaction} by pressure increase, the fact that
denser sand can sustain more compression before getting
unjammed, before elastic solutions become unstable: See
the dotted line of Fig~\ref{fig1}-(a), depicting a
well-known empirical formula from soil
mechanics~\cite{elaPla-1,elaPla-2,elaPla-3},
$e=e_0-\Lambda \ln P$. Referred to as the {\em virgin}
(or {\em primary) consolidation line}, it represents the
boundary that sand (at rest) will not cross when
compressed. Instead, it will collapse, becoming more
compact, with a smaller $e$, close to or at the curve,
but not beyond. (Note the dotted line does not appear to
cut the $e$-axis, as it should at $\rho_{\ell p}$ -- this
is where sand becomes instable for any pressure. The
discrepancy may derive from difficulties of making
reliable measurements close to $\rho_{\ell p}$.)

This behavior is a natural consequence of higher-order
strain terms such as the next order ones
($\zeta_1,\zeta_2>0$),
\begin{equation}\label{fe11}
-(\zeta_1\Delta^3 +\zeta_2\Delta u_s^2),
\end{equation}
which need to be added to $w_1$ as given by
Eqs~(\ref{fe3},\ref{fe10}). Consider first pure
compression, $u_s^2=0$. For small $\Delta$, the term
$-\zeta_1\Delta^3$ is negligible, and $w_1$ remains
convex. But if $\Delta$ is large enough, its negative
second derivative will turn $w_1$ concave, making any
elastic solution impossible. The value of $\Delta$ at
which this happens, grows with $\cal B$ -- a larger
third-order term is needed for a larger $\cal B$. Now,
$\cal B$ is smallest at $\rho=\rho_{\ell p}$, grows
monotonically with $\rho$, and diverges at
$\rho_{cp}$. As a result, the instability line cuts
the $e$-axis at $\rho_{\ell p}$, veers towards larger
$\Delta$ (or larger $P$) at higher density , and heads
for infinity at $\rho_{cp}$, see the thin line
depicted as ``simple model" in Fig~\ref{fig1}-(a),
drawn with a constant $\zeta_1=24500$ MPa. (It is of
course possible, employing a density-dependent
$\zeta_1$, to improve the agreement to the dotted
line.) In Fig~\ref{fig1}-(b), the point of maximal
pressure for a given void ratio $e$ is located at
where the $P$-axis is being cut by the associated
curve. If the term $\sim\Delta u_s^2$ did not exist,
these curves would be vertical lines. The presence of
$\sim\Delta u_s^2$ reduces the value of $\Delta$ (or
$P$) for growing $u_s$ (or $\sigma_s$), bending the
lines to the left.

Although qualitative figures of these curves that are
frequently referred to as {\em caps} abound in
textbooks~\cite{elaPla-1,elaPla-2,elaPla-3}, we did not
find enough quantitative data, especially not a generally
accepted empirical expression, that we could have
compared our results to. Presumably, it is not easy to
observe caps in dry sand. Given this lack of reliable
data, we decided against the expansion, Eq~(\ref{fe11}),
and opted for a flexible ``cap function," $\cal C$ of
Eq~(\ref{fe10}), capable of accounting for any possible
cap-like unjamming transitions,
\begin{eqnarray}\label{fe12a}
2{\cal C}=1+\tanh [(\Delta _0-\Delta)/\Delta _1],
\quad \text{where}\quad\\ \Delta _0=k_1\rho
-k_2u_s^2-k_3=k_1^{\prime }
/(e+1)-k_2u_s^2-k_3.\nonumber
\end{eqnarray}
With ${\cal C}\approx1$ for  $\Delta\ll\Delta_0$, and
${\cal C}\approx0$ for $\Delta\gg\Delta_0$, the cap
function is constructed to be relevant only in a
narrow neighborhood around $\Delta_0$, for
$|\Delta-\Delta_0|\lesssim\Delta_1\approx 10^{-4}$,
such that the energy's convexity is destroyed around
$\Delta_0$. Taking $k_1, k_2, k_3$ as constant,
$\Delta_0$ grows with the density and falls with
$u_s^2$, giving rise to the typical appearance
reproduced in Fig~\ref{fig1}.

Together, Eqs~(\ref{fe3},\ref{fe10},\ref{fe12a}) give
the energy density $w_1$, appropriate for cohesionless
granular materials at $T_g=0$. There are two
contributions to the  pressure, $P=P_1+P_\Delta$,
where $P_1\equiv\rho(\partial w_1/\partial\rho)-w_1$
from Eq~(\ref{fe4}), and $\pi_{ij}=-\partial
w_1/\partial u_{ij}\equiv P_\Delta\delta
_{ij}-\sigma_s u_{ij}^0/u_s$. Because we still take
$\Delta$ to be a small quantity, $P_1\sim\Delta^{2.5}$
may be neglected. (Similarly, terms such as
$\pi_{ik}u_{jk}\sim\Delta^{2.5}$ from
Eq~(\ref{7y-GSH}) below are also negligible.) So the
stress is simply $\pi_{ik}$, with pressure and shear
stress given as
\begin{eqnarray}
P_\Delta &=&{\cal B}\sqrt{\Delta }(\Delta
+{\textstyle\frac 3{10}}u_s^2/\Delta) -w_1{\cal
C}^*/\Delta _1, \label{fe13}
\\ \sigma _s &=&{\textstyle\frac 65}{\cal B}
\sqrt{\Delta} u_s- 2k_2u_sw_1 {\cal C}^*/\Delta_1,
\label{fe14}
\end{eqnarray}
where ${\cal C}^*\equiv1-\tanh [(\Delta _0-\Delta
)/\Delta _1]$, hence ${\cal C}^*\to0$ away from the
cap. (The terms of higher order in $\Delta$ are kept
in ${\cal C}^*$, because $\Delta_1$ is small. This is
how we make $\cal C$ a function relevant for
$\Delta\approx\Delta_0$, not $\Delta\to0$.)

Stability is given only if the energy $w_1$ is convex
with respect to its seven variables, $\rho, \Delta,
u_{ij}^0$. As linear transformations do not alter the
convexity property of any function, we may take the
energy as $w_7(\rho, \Delta, x_{1-5})$ where
$x_1\equiv\sqrt{2 }u_{xy}$, $x_2\equiv\sqrt{2}u_{xz}$,
$x_3\equiv\sqrt{2}u_{yz}$,
$x_4\equiv(u_{xx}-u_{zz})/\sqrt{2}$,
$x_5\equiv(u_{xx}-2u_{yy}+u_{zz})/\sqrt{6}$. The
characteristic polynomial $N_7$ of the Hessian matrix of
$w_7$ is $N_7=(\lambda-u_s^{-1}\partial w_1/\partial
u_s)^4 N_3$, with $N_3$ the characteristic polynomial of
$w_1(\Delta,u_s,\rho)$. Since $u_s^{-1}\partial
w_1/\partial u_s$ is always positive, it is sufficient to
consider $w_1(\Delta,u_s,\rho)$. Requiring $N_3$ to have
only positive eigenvalues defines the stable region in
the strain space, spanned by $\Delta,u_s,e$. Using
Eqs~(\ref{fe13},\ref{fe14}), we may convert this into one
in the stress space, spanned by $P,\sigma_s,e$. The
result, obtained numerically, is the yield surface
plotted in Fig~\ref{fig1}.

\subsection{Pressure Contribution From Agitated Grains\label{P_T}}
Agitated grains are known to exert a pressure in granular
liquid. Using the model of ideal gas (better:
non-interacting atoms with excluded volumes), with
$w_2\sim{\rho T_g}$ denoting the energy density of
agitated grains, the pressure expression,
\begin{equation}\label{fe5}
P_T(\rho, T_g)\sim{w_2}/({1-\rho/\rho_{cp}}),
\end{equation}
was employed and found to account realistically for the
behavior of granular liquid sandwiched between two
cylinders rotating at different
velocities~\cite{Lub-1,Lub-2,Lub-3,Lub-4}.

In ideal gas, both the energy density $w$ and pressure
$P$ are proportional to the temperature $T$. As a
consequence, the entropy is $s\sim\ln T$, and diverges
for $T\to0$. (The free energy has a contribution $\sim
T\ln T$ that vanishes for $T\to 0$.) As quantum effects
become important long before $T$ vanishes, the unphysical
feature of a diverging entropy is inconsequential for
ideal gases. Yet this would be a highly relevant defect
for granular solids, for which important physics occurs
at or around $\bar T_g=0$. This is the reason ideal gas
is not an appropriate model for granular solids. The
considerations of section~\ref{Tg} show that
$w_2,f_2\sim\bar T_g^2$ close to $\bar T_g=0$ -- implying
a pressure contribution, $P_2=\rho(\partial
f_2/\partial\rho)-f_2\sim\bar T_g^2$, see Eq~(\ref{fe4}).
Note  first that $P_2\sim w_2$ is retained, and second
that because $P_0=0$, $P_1\approx0$, we have
$P_T\equiv\sum P_i\approx P_2$.

Unfortunately, the density dependence of
Eq~(\ref{fe5}) also poses a problem, as it implies a
free energy
$f_2=b_0\rho\ln(1-\rho/\rho_{cp})(-T_g^2/2)$ and a
granular entropy, $s_g=-\partial f_2/\partial T_g=
b_0\rho\ln(1-\rho/\rho_{cp})\, T_g$, both diverging
for $\rho\to\rho_{cp}$. We therefore take $f_2$ to be
given as in Eq~(\ref{fe2}), with a positive but small
$a$. The resulting entropy is physically acceptable,
and the pressure is easily rendered numerically
indistinguishable from  Eq~(\ref{fe5}),
\begin{eqnarray}
P_T&=&P_2= \frac{\rho}{2\rho_{cp}}\frac{a\, \rho \,b_0
\bar T_g^2}{(1-\rho/\rho_{cp})^{1-a}},\label{fe6}
 \\ s_g&=&-\frac{\partial f_2}{\partial \bar T_g}=
\rho\,b_0 \bar
T_g\left(1-\frac\rho{\rho_{cp}}\right)^{a}.\label{fe7}
\end{eqnarray}

\begin{figure}[tb]
\begin{center}
\includegraphics[scale=0.6]{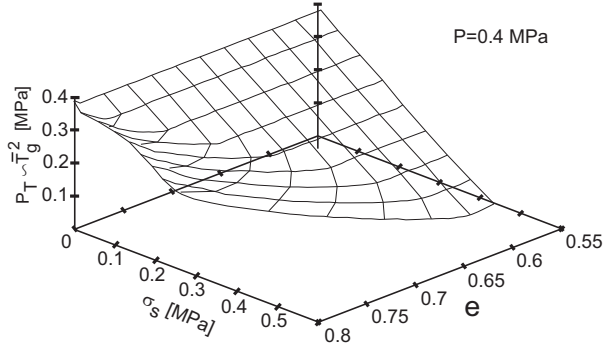}
\end{center}
\caption{Jamming transition as a function of $e,
\sigma_s$ and $P_T\sim T_g^2$, for $P_\Delta=0.4$ MPa.
Values of model parameters are the same as those in
FIG.~1.} \label{jamming}
\end{figure}

As the total pressure is now $P=P_T+P_\Delta$, cf.
Eq~(\ref{fe13}), the jamming transition discussed
above is modified. For instance, the yield condition
of Eq~(\ref{1-2a}), with $\xi=5/3$, now reads
\begin{equation}\label{fe15}
\frac{\pi_s}{P_\Delta}=
\frac{\pi_s}{P-P_T}\le\sqrt{\frac65},
\end{equation}
implying a smaller maximal $\pi_s$ for given $P$. On the
other hand, the maximal value for the void ratio $e$ is
larger when $P_T$ is present: Any given $e$ has a maximal
elastic compression $\Delta$ that will not sustain a
larger $e$. But if $P$ is fixed and $T_g$ is finite, the
elastic compression $\Delta$  will be appropriately
smaller to sustain a larger $e$. This behavior is
depicted in Fig~\ref{jamming}.

The jamming transition, from elastic solid to liquid,
is of course no longer completely sharp at a finite
$T_g$, because $T_g$ turns the elastic body into a
transiently elastic one for all values of stress and
density. Nevertheless, there is a huge quantitative
difference between catastrophic unjamming and the
gradual process of stress relaxation. A sand pile may
slowly degrade, relaxing toward the flat surface. But
when turning on $T_g$ violates Eq~(\ref{fe15}), sudden
events such as liquefaction happen. ($P_T$ may be
substituted by the pore pressure to account for a
similar collapse, if the soil is filled with water.)
The frequently reported phenomenon of a primary
earthquake emitting elastic waves that trigger
earthquakes elsewhere~\cite{Jia2}, may well be
connected to Eq~(\ref{fe15}): $T_g$ as given by
Eq~(\ref{2-21}) accompanies elastic waves. It may be
sufficiently large to violate Eq~(\ref{fe15}) if
stability was precarious.

\subsection{The Edwards Entropy}

It is useful, with the free energy obtained in this
chapter in mind, to revisit the starting points of
Granular Statistical Mechanics ({\sc gsm}), especially
the Edwards entropy~\cite{Edw}. Taking the entropy
$S(W,V)$ as a function of the energy $W$ and volume $V$,
or ${\rm d}S=(1/T){\rm d}W+(P/T){\rm d}V$, it argues that
{\it a mechanically stable agglomerate of infinitely
rigid grains at rest} has, irrespective of its volume,
vanishing energy, $W\equiv0$, ${\rm d}W=0$. The physics
is clear: However we package these rigid grains that
neither attract nor repel each other, the energy remains
zero. Therefore, ${\rm d}S=(P/T){\rm d}V$, or ${\rm
d}V=(T/P){\rm d}S\equiv X{\rm d}S$. This is the starting
expression of {\sc gsm}, and $X$ is considered the
relevant quantity characterizing granular media at rest.
The entropy $S$ is obtained by counting the number of
possibilities to package grains for a given volume,
taking it to be $e^S$. And because a stable agglomerate
is stuck in one single configuration, some tapping (or a
similar disturbance) is taken to be needed to enable the
system to explore the phase space.

In {\sc gsh}, the grains are neither infinitely rigid,
nor generally at rest. An elastic and a $T_g$-dependent
energy contribution, denoted respectively as $f_1$ and
$f_2$, see Eq~(\ref{fe1}), account for them. {\sc gsh}
also possesses a $T_g$-switch that determines whether the
system's behavior is solid- or liquid-like. This is
clearly the generalization of phase space exploration
enabled by tapping. That grains neither attract nor repel
each other is accounted for by the stress vanishing for
$T_g,u_{ij}\to0$: In this limit, in which $f_1,f_2=0$ and
only $f_0\sim\rho$ finite, there is no term in the energy
that depends nonlinearly on the density $\rho$, hence
$\sigma_{ij}=\partial (f_0/\rho)/\partial
(1/\rho)\delta_{ij}=0$.


Given this comparison, it is natural to ask whether {\sc
gsm} is a legitimate limit of {\sc gsh}. The answer is
probably no, as both appear conceptually at odds in two
points, the first more direct, the second quite
fundamental: (1)~Because of the Hertz-like contact
between grains, very little material is being deformed at
first contact, and the compressibility diverges at
vanishing compression. This is a geometric fact
independent of how rigid the bulk material is. Infinite
rigidity is therefore not a realistic limit for sand.
(2)~In considering the entropy, one must not forget that
the number of possibilities to package grains for a given
volume is vastly overwhelmed by the much more numerous
configurations of the inner granular degrees of freedom.
Maximal entropy $S$ for given energy therefore
realistically implies minimal macroscopic energy, such
that a maximally possible amount of energy is in $S$ (or
heat), equally distributed among the numerous inner
granular degrees of freedom. Maximal number of
possibilities to package grains for a given volume is a
fairly different criterion.

\section{Granular Hydrodynamic Theory\label{GHT1}}
\subsection{Derivation\label{GHT}}
We take the conserved energy $w(s,s_g,\rho,g_i,u_{ij})$
of granular media to depend on entropy $s$, granular
entropy $s_g$, density $\rho$, momentum density $g_i$,
and the elastic strain $u_{ij}$. Defining the conjugate
variables as $T\equiv\partial w/\partial s$, $\bar
T_g\equiv T_g-T\equiv\partial w/\partial s_g$ [see
Eq~(\ref{2-13})], $\mu-v^2/2\equiv\partial
w/\partial\rho$ [see Eq~(\ref{2-3a})], $v_i\equiv\partial
w/\partial g_i=g_i/\rho$ [see Eq~(\ref{2-1})],
$\pi_{ij}\equiv-\partial w/\partial u_{ij}$, we write
\begin{eqnarray}
\label{20-GSH} {\rm d}w = T{\rm d} s +\bar T_g{\rm d}s_g
+ (\mu-v^2/2) {\rm d}\rho\\\nonumber + v_{i} {\rm d}
 g_{i} - \pi_{ij} {\rm d} u_{ij}.
\end{eqnarray}
The equations of motion for the energy and its variables
are
\begin{eqnarray}
\label{2y-GSH} \partial_t w+\nabla_iQ_i=0, \quad
\partial_t \rho+\nabla_i(\rho
v_i)=0,\qquad
\\
\partial_t g_i+\nabla_j
(\sigma_{ij}+g_iv_j)=0,\qquad\qquad\label{3yA-GSH}
\\ \label{3y-GSH}
\partial_t
s+\nabla_if_i=R/T,\quad\!\! \partial_t
s_g+\nabla_iF_i=R_g/T_G,
\\
\label{4y-GSH} {\rm d}_t
u_{ij}-v_{ij}-X_{ij}=\qquad\qquad\qquad\qquad
\\\nonumber-[(u_{ik}\nabla_{j}v_k+\nabla_iy_j/2)
+(i\leftrightarrow j)].
\end{eqnarray}
The first three equations are conservation laws, with the
fluxes $Q_i$ and $\sigma_{ij}$ as yet unknown, to be
determined in this section. The next two are the balance
equation for the two entropies, the form of which are
already given, in
Eqs~(\ref{2-17},\ref{2-18a},\ref{2-19},\ref{2-20}).
Nevertheless, to see that they indeed fit the constraints
required by energy and momentum conservation, we
designate the currents as $f_i=sv_i-f_i^D$,
$F_i=s_gv_i-F^D_i$, leaving $f_i^D, F^D_i, R, R_g$
unspecified. The last is the equation of motion for the
elastic strain field, as discussed in
section~\ref{elaPla}, with $y_i, X_{ij}$ the unknown
fluxes to be determined here. Next, we introduce
$\sigma^D_{ij}+\Sigma^D_{ij}$, as
\begin{eqnarray}\label{7y-GSH}
\sigma_{ij}\equiv(-\tilde f+\mu\rho)
\delta_{ij}-(\sigma^D_{ij}+\Sigma^D_{ij})\\
+\pi_{ij}-\pi_{ik}u_{jk}-\pi_{jk}u_{ik}, \nonumber
\end{eqnarray}
where $\tilde f\equiv w_0-Ts-\bar T_gs_g$, as in
Eq~(\ref{2-14},\ref{fe4}). This is simply a definition of
$\sigma^D_{ij}+\Sigma^D_{ij}$, which transfer our task
from determining $\sigma_{ij}$ to finding the new
quantity. This simplifies our task, notationally, of
finding the form of $\sigma_{ij}$, it does not in anyway
prejudice it.

Differentiating the energy, ${\partial_ t}w = T{\partial_
t} s +\bar T_g{\partial_ t} s_g+ (\mu-v^2/2) {\partial_
t}\rho + v_{i}{\partial_ t} g_{i} - \pi_{ij} {\partial_
t} u_{ij}$, see Eq~(\ref{20-GSH}), then inserting
Eqs~(\ref{2y-GSH},\ref{3yA-GSH},\ref{3y-GSH},\ref{4y-GSH})
into it, employing relations such as $\bar T_g\partial_t
s_g=\bar T_gR_g/T_G+v_ks_g\nabla_k\bar T_g -\nabla_k(\bar
T_gs_gv_k)$,
we obtain
\begin{eqnarray}\label{xx1}
\nabla_iQ_i=\nabla_i(Tf_i+\bar T_gF_i+\mu\rho v_i
+v_j\sigma_{ij}-y_j\pi_{ij}) \qquad
\\\nonumber
-R+f_i^D\nabla_iT +\sigma_{ij}^Dv_{ij}
+y_i\nabla_j\pi_{ij}+ X_{ij}\pi_{ij}+\gamma\bar
T_g^2\\\nonumber
-R_g+\Sigma_{ij}^Dv_{ij}+F_i^D\nabla_i\bar T_g
-\gamma\bar T_g^2
\end{eqnarray}
This is a useful result, which shows one can rewrite
${\partial_ t}w$ as the divergence of something (first
line), plus something (second and third line) that
vanishes in equilibrium -- see section~\ref{granular
equilibria} why $\nabla_iT, v_{ij}, \pi_{ij},
\nabla_j\pi_{ij}$ and $T_G$ vanish. We take the first
line to yield the energy flux, $Q_i$, and the next two
lines to vanish independently,
\begin{eqnarray}\label{9y-GSH}
Q_i&=&Tf_i+\bar T_gF_i+\mu\rho v_i
+v_j\sigma_{ij}-y_j\pi_{ij}, \\
R&=&f_i^D\nabla_iT+\sigma_{ij}^Dv_{ij}
+y_i\nabla_j\pi_{ij} +X_{ij}\pi_{ij}+\gamma \bar T_g^2,\\
R_g&=&\Sigma_{ij}^Dv_{ij} +F_i^D\nabla_i\bar T_g -\gamma
\bar T_g^2.
\end{eqnarray}
Comparing $R,R_g$ with Eqs~(\ref{2-18a},\ref{2-20}),
the currents are found as
\begin{eqnarray}
\label{12y-GSHy}f^D_i=\kappa\nabla_iT,\quad
\sigma_{ij}^D=\zeta v_{\ell\ell}\delta_{ij}+\eta
v^0_{ij}+\alpha\pi_{ij},
\\\nonumber%
F^D_i=\kappa_g\nabla_i\bar T_g,\quad\quad
\Sigma_{ij}^D=\zeta_g v_{\ell\ell}\delta_{ij}+\eta_g
v^0_{ij},
\\\nonumber
y_i=\beta^{P}\nabla_j\pi_{ij},\,\, X_{ij}=\beta\pi_{ij}^0
+\beta_1\delta_{ij}\pi_{\ell\ell}-\alpha v_{ij}.
\end{eqnarray}
(It is an assumption to take $F_i^D\nabla_i\bar T_g$ as
part of $R_g$ rather than $R$.) The two terms preceded by
$\alpha$ contribute $\pm\alpha\pi_{ij}v_{ij}$ to $R$,
respectively, hence cancel each other and are compatible
with Eq~(\ref{2-18a}). (More such pairs of terms,
mutually canceling or contributing in equal parts, are
possible. They have been excluded as a simplification. In
the language of the Onsager force-flux relation, the
above fluxes possess only diagonal elements, with the
exception of the reactive, off-diagonal terms
$\sim\alpha$.) Defining two relaxation times,
\begin{equation}\label{GSH1}
\frac1\tau\equiv2\beta{\cal A}\sqrt\Delta, \quad
\frac1{\tau_1}\equiv3\beta_1\sqrt\Delta\left({\cal
B}+\frac{{\cal A}u_s^2}{2\Delta^2}\right).
\end{equation}
the last of Eqs~(\ref{12y-GSHy}) may be written as
\begin{equation}\label{12y-GSH}
 X_{ij} =\Delta\,\delta_{ij}/{\tau_1}-{u
_{ij}^0}/\tau-\alpha v_{ij}.
\end{equation}
To ensure permanent elasticity in granular statics, we
must in addition require
\begin{equation}\label{2GSH}
X_{ij}\to0\quad \text{for}\quad T_g\to0.
\end{equation}
This completes the derivation of {\sc gsh}. Given $f_i^D,
F_i^D, \sigma_{ij}^D, \Sigma_{ij}^D, y_i, X_{ij}$, the
structure of all currents in the set of equation,
Eqs~(\ref{2y-GSH},\ref{3yA-GSH},\ref{3y-GSH},\ref{4y-GSH}),
are known. The question that remains is whether these
expressions are unique. For simpler hydrodynamic
theories, such as for isotropic liquid, nematic liquid
crystal, or elastic solid, this procedure (frequently
referred to as the {\em standard procedure}) is easily
shown to be unique, because one can convince oneself that
as long as the energy $w$ remains unspecified, there is
only one way to write the time derivative of the energy
$\partial_tw$ as the sum of a divergence and a series of
expressions that vanish in equilibrium. In the present
case, with two levels of entropy productions, one of
which controls the switch between permanent and transient
elasticity, the hydrodynamic theory is singularly
intricate, and peripheral ambiguity remains.
Nevertheless, displaying energy and momentum conservation
explicitly, and reducing to liquid and solid
hydrodynamics in the proper limits, the given set of
equations is certainly a viable and consistent theory.

A more formal way of obtaining the fluxes of
Eqs~(\ref{12y-GSHy}) is to define the flux and force
vectors as $\vec Z=(f^D_i, y_i, \sigma^D_{ij}, X_{ij})$,
$\vec Z_g=(F^D_i, \Sigma^D_{ij})$, $\vec Y=(\nabla_iT,
\nabla_j\pi_{ij}, v_{ij}, \pi_{ij})$, $\vec
Y_g=(\nabla_i\bar T_g, v_{ij})$. And because $R=\vec
Z\cdot\vec Y$, $R_g=\vec Z_g\cdot\vec Y_g$, the Onsager
force-flux relations are given as
\begin{equation}
\vec Z=\hat c\cdot\vec Y, \quad \vec Z_g=\hat
c_g\cdot\vec Y_g.
\end{equation}
The transport matrices, $\hat c$ and  $\hat c_g$, have
positive diagonal elements and off-diagonal ones that
satisfy the Onsager reciprocity relation. Our example
above has only diagonal elements, with the single
exception of the reactive, off-diagonal terms
$\sim\alpha$.

\subsection{Results}

Collecting the terms derived above, in section~\ref{GHT},
the equations of {\sc gsh}, with $\sigma_{ij}$ valid to
lowest order in strain, are
\begin{eqnarray}
\partial_t \rho+\nabla_i(\rho
v_i)=0,\qquad\qquad\quad\label{14y-GSH}
\\\nonumber {\rm d}_t
u_{ij}=(1-\alpha)v_{ij}-{u _{ij}^0}/\tau
-{u_{\ell\ell}\,\delta_{ij}}/{\tau_1}\qquad
\\\label{15y-GSH} -(u_{ik}\nabla_jv_k+
\nabla_i[\beta^P\nabla_k\pi_{jk}/2])-(i\leftrightarrow
j),
\\\nonumber \sigma_{ij}=(1-\alpha)\pi_{ij}
-\pi_{ik}u_{jk}-\pi_{jk}u_{ik}\qquad\qquad\quad
\\\label{16y-GSH}
+(\mu\rho-\tilde f)\delta_{ij} -(\zeta+\zeta_g)
v_{\ell\ell}\delta_{ij}-(\eta+\eta_g) v^0_{ij},
\\\nonumber T_g[\partial_t
s_g+\nabla_i(s_gv_i-\kappa_g\nabla_i\bar T_g)]=R_g=
\qquad\quad
\\\label{8y-GSH}
 \zeta_g v_{\ell\ell}^2+\eta_g
v^0_{ij}v^0_{ij}+\kappa_g(\nabla_i\bar T_g)^2-\gamma\bar
T_g^2,\,\,\,
\\\nonumber
T[\partial_t s+\nabla_i(sv_i-\kappa\nabla_iT)]= \zeta
v_{\ell\ell}^2+\eta v^0_{ij}v^0_{ij}+\gamma\bar T_g^2
\\\label{8ya-GSH}
+\kappa(\nabla_iT)^2+\beta^P(\nabla_j\pi_{ij})^2
+\beta\pi^0_{ij}\pi^0_{ij}+\beta_1\pi_{\ell\ell}^2.\,\,
\end{eqnarray}
Given in terms of the variables: ($s$, $s_g$, $\rho$,
$g_i$, $u_{ij}$), conjugate variables ($T$, $\bar T_g$,
$\mu$, $v_i$, $\pi_{ij}$), and 11 transport coefficients,
($\alpha$, $\tau$, $\tau_1$, $\zeta$, $\zeta_g$, $\eta$,
$\eta_g$, $\gamma$, $\beta^P$, $\kappa$, $\kappa_g$),
these equations are valid irrespective of the functional
form of the energy $w$ and the transport coefficients.
Therefore, they only provide a hydrodynamic structure, a
framework into which different concrete theories fit.
This circumstance, though also true for Newtonian fluids,
is not as relevant there, because static susceptibilities
(such as the compressibility or specific heat) and
transport coefficients may frequently be approximated as
constant. So the structure alone already possesses
considerable predicting power. This is not true for
granular media, which typically possess more involved
functional dependence -- especially concerning the $\bar
T_g\to0$ limit, which does not have a counter part in
other systems. This is one of the less recognized
reasons, we believe, underlying the complexity of
granular systems.

In section~\ref{GraFE}, a free energy was proposed that
we are confident should be fairly realistic. The
situation with respect to the 11 transport coefficients
are less settled, and in need of much future work, though
a few limits are clear from the onset: First, a simple,
analytic way to assure the elastic limit for $\bar T_g=0$
and satisfy the requirement of Eq~(\ref{2GSH}) is given
by
\begin{equation}\label{Re-app3}
1/\tau=\lambda\bar T_g,\quad 1/\tau_1=\lambda_1\bar T_g,
\end{equation}
which, as we shall see next, gives rise to the same
dynamic structure as hypoplasticity. The density
dependence is more subtle, hence harder and less urgent
to determine. However, it seems plausible that $\lambda,
\lambda_1$ should decrease for growing density, and
especially the compressional relaxation should stop being
operative at the random close packing density
$\rho_{cp}$. To account for this, the simplest dependence
would be
\begin{equation}
  \lambda_1\sim (\rho-\rho_{cp}).
\end{equation}

The coefficient $\alpha$  needs to vanish in the elastic
limit, for $\bar T_g\to0$, and be constant in the
hypoplastic one, when $\bar T_g$ is moderately large: We
have $\sigma_{ij}=\pi_{ij}$ in the elastic regime, and
$\sigma_{ij}=(1-\alpha) \pi_{ij}+\cdots$ with
$1-\alpha\approx0.2$ in the hypoplastic one, implying
sand is much softer here -- same strain, yet stress is
smaller by a factor of about five. The behavior of
$\alpha$ is probably the result of granular agitation
disrupting force chains. They are all intact in the
elastic limit, making the system comparatively stiff. A
finite $\bar T_g$ breaks up the chains, and when most of
chains are destroyed, the remaining ones become essential
in the sense that their disruption leads to local
collapse, which in turn immediately repair the chains by
some rearrangement. This is why $\alpha$ saturates and
becomes constant.

Finally, as long as Eq~(\ref{xx2-222}) holds, the rate
independence it entails would prevent the propagation of
sound and elastic waves: Because both the elastic and the
plastic part are linear in the velocity, and of the same
order in the wave vector $q$, sound damping is comparable
to sound velocity, and wave propagation could at most
persist for a few periods. We therefore expand
$\gamma,\eta_g$ in $\bar T_g$, as
\begin{equation}
\gamma=\gamma_0+\gamma_1\bar T_g,\quad \eta_g=\eta_1\bar
T_g,
\end{equation}
assuming $\eta_g$ lacks a constant term, because viscous
dissipation occurs directly via $\eta$ for $\bar
T_g\to0$, see Eq~(\ref{12y-GSHy}). Inserting these
expression into Eq~(\ref{2-21}) for a quick, qualitative
estimate, we find $\bar T_g\sim{v_{ij}v_{ij}}\equiv
v_s^2$ for $\gamma_0\gg\gamma_1\bar T_g$, and $\bar
T_g\sim v_s$ for $\gamma_0\ll\gamma_1\bar T_g$. The first
regime is essentially elastic, because the relaxation
term, $u_{ij}/\tau\sim u_{ij}\bar T_g\sim u_{ij}v_s^2$,
is of second order and small. This ensures the
propagation of sound modes. In the second regime, the
same term, $u_{ij}/\tau\sim u_{ij}\bar T_g\sim
u_{ij}v_s$, is of first order and rather more prominent,
giving rise to the hypoplastic behavior discussed in the
next section .

\section{The Hypoplastic Regime\label{hypo}}

{\em Hypoplasticity}~\cite{Kolym-1}, a modern,
well-verified, yet comparatively simple theory of soil
mechanics, models solid dynamics as realistically as the
best of the elasto-plastic theories. Its starting point
is the rate-independent constitutive relation,
\begin{equation}
\partial_t{\sigma }_{ij}
=H_{ijk\ell}\,\,v_{k\ell}+\Lambda
_{ij}\sqrt{v^0_{ij}v^0_{ij}+\epsilon\,
v_{\ell\ell}^2}, \label{hp1}
\end{equation}
where the coefficients $H_{ijk\ell},\Lambda
_{ij},\epsilon$ are functions of $\sigma_{ij},\rho$,
specified using experimental data mainly from triaxial
apparatus. (Rate-independence means $\partial_t{\sigma
}_{ij}$ is linearly proportional to the magnitude of the
velocity, such that the change in stress remains the same
for given displacement  irrespective how fast the change
is applied, a well verified observation in both the
elastic and hypoplastic regime.) Great efforts are
invested in finding accurate expressions for the
coefficients, of which a recent set~\cite{Kolym-1} is
$\epsilon=1/3$,
\begin{eqnarray}\label{hp2}
H_{ijk\ell} &=&f\left( F^2\delta _{ik}\delta
_{j\ell}+a^2\sigma _{ij}\sigma _{k\ell}/\sigma
_{nn}^2\right) \text{,}
\\ \Lambda _{ij} &=&aff_dF\left( \sigma
_{ij}+\sigma _{ij}^0\right) /\sigma _{nn}, \label{hp3}
\end{eqnarray}
where $a=2.76$, $h_s=1600$~MPa,$\ e_d=0.44e_i$,
$e_c=0.85e_i$, $e_i^{-1}=\exp \left( \sigma
_{\ell\ell}/h_s\right) ^{0.19}$, and
\begin{eqnarray*} f_d =\left(
\frac{e-e_d}{e_c-e_d}\right) ^{0.25}\!\!\!\!\!\!,\quad
f=-\frac{ 8.7h_s\left( 1+e_i\right) }{3\left(
\sigma_s/\sigma_{\ell\ell}+1\right) e}\left( \frac{\sigma
_{\ell\ell}}{ h_s}\right) ^{0.81}\!\!\!\!\!\!,
\\ \nonumber F =\sqrt{\frac{3\sigma _s^2}{8\sigma _{\ell\ell}
^2}+\frac{2\sigma _s^2\sigma _{\ell\ell}-3\sigma
_s^4/\sigma _{\ell\ell}}{2\sigma _s^2\sigma
_{\ell\ell}-6\sigma _{ij}^0\sigma _{j\ell}^0\sigma _{\ell
i}^0}}-\sqrt{\frac 38}\frac{\sigma _s}{\sigma
_{\ell\ell}}.
\end{eqnarray*}

{\sc gsh}, as derived above, reduces to Eq~(\ref{hp1})
for a stationary $T_g$, with $H_{ijlk},\Lambda
_{ij},\epsilon$ given in terms of
$M_{ijk\ell}\equiv-\partial^2 w/\partial
u_{ij}\partial u_{k\ell}$ and four scalars that are
combinations of transport coefficients. We assume
uniformity and stationarity, with especially
$\nabla_i\bar T_g, \nabla_j\pi_{ij}, \partial_t
v_i=0$, and only include the lowest order terms in the
strain $u_{ij}$. We also take $\alpha,\eta_g,\zeta_g$
as constants, and neglect $P_T$ from Eq~(\ref{fe6}),
the pressure relevant in granular liquid, assuming
$T_g$ is too small for the given velocity. It is then
quite easy to evaluate $\partial_t \sigma_{ij}$
employing Eqs~(\ref{15y-GSH},\ref{16y-GSH}),
\begin{eqnarray}\nonumber
{\partial_t}\sigma_{ij}=(1-\alpha){\partial_t}
\pi_{ij}=(1-\alpha)M_{ijk\ell} {\partial_t} u_{k\ell}=\\
{\textstyle(1-\alpha) M_{ijk\ell} [(1-\alpha)v_{k\ell}-
u^0_{k\ell}/\tau-\delta_{k\ell}u_{mm}/{\tau_1}]}.
\label{gsh4}\end{eqnarray}
Clearly, given Eqs~(\ref{2-21},\ref{Re-app3}), this
expression already has the structure of Eq~(\ref{hp1})
that {\em Hypoplasticity} postulates. And the
coefficients are
\begin{eqnarray}\label{hpA}
H_{ijk\ell}=(1-\alpha)^2 M_{ijk\ell},\qquad
\epsilon=\zeta _g/\eta _g,\\
\Lambda_{ij}=(1-\alpha)M_{ijk\ell}
[(\tau/\tau_1)\Delta\delta_{k\ell}-u_{k\ell}^0]\lambda
{\sqrt{\eta_g/\gamma}}. \label{hpB}\end{eqnarray}
{\sc hpm} has 43 free parameters (36+6+1 for
$H_{ijk\ell}, \Lambda_{ij}, \epsilon$), all functions of
the stress and density. Expressed as here, the stress and
density dependence are essentially determined by
$M_{ijk\ell}$ that is a known quantity~\cite{ge-1,ge-2}.
For the four free constants, we take
\begin{eqnarray}\label{last}
1-\alpha=0.22,\qquad\tau/{\tau_1}=0.09,
\\\nonumber
\frac{\zeta _g}{\eta_g}=0.33,\,
\sqrt{\frac{\eta_g}{\gamma}}=
\sqrt{\frac{\eta_1}{\gamma_1}}=\frac{114}\lambda,
\end{eqnarray}
to be realistic choices, as these numbers yield
satisfactory agreement with {\em Hypoplasticity}.
Their significance are: $\zeta _g/\eta _g=0.33$
implies shear flows are three times as effective in
creating $T_g$ as compressional flows.
$\tau/\tau_1=0.09$ means, plausibly, that the
relaxation rate of shear stress is ten times higher
than that of pressure. The factor $(1-\alpha)^2$
accounts for an overall softening of the static
compliance tensor $M_{ij\ell k}$. Finally, $\lambda$
controls the stress relaxation rate for given $T_g$,
and ${\textstyle\sqrt{\eta_1/\gamma_1}}$ how well
shear flow excites $T_g$. Together,
$\lambda{\textstyle\sqrt {\eta_g/\gamma}}=114$
determines the relative weight of plastic versus
reactive response: The term in Eq~(\ref{hp1}) preceded
by $H_{ijk\ell}$ is the reversible, elastic response,
the second term preceded by $\Lambda_{ij}$ comes from
stress relaxation, is dissipative, irreversible and
plastic. For small strain, $\Delta,u_s\to0$, the
elastic part is dominant,
$|H_{ijk\ell}|\gg|\Lambda_{ij}|$. But $|\Lambda_{ij}|/
|H_{ijk\ell}| \sim|u_{k\ell}^0|\cdot114/(1-\alpha)$ is
of order unity for $|u_{ij}|$ around $10^{-3}$.

Although the functions of Eqs~(\ref{hpA},\ref{hpB})
appear rather different from that of
Eqs~(\ref{hp2},\ref{hp3}), the stress-strain
increments are quite similar, as the comparison
in~\cite{JL3} shows. Moreover, the residual
discrepancies may be eliminated by discarding the
simplifying assumption of constant transport
coefficients, independent of the stress. This
agreement provides valuable insights into the physics
of {\em Hypoplasticity}, showing why it works, what
its range of validity is, and how it may be
generalized. And it conversely also verifies {\sc
gsh}.

\section{Conclusion\label{conclu}}

The success of {\em Granular Elasticity}, the theory
we employ to account for static stress distribution in
granular media, is mainly due to the fact that the
information on the plastic strain is quite irrelevant
there. This is no longer true in granular dynamics,
when the system is being deformed -- sheared,
compressed or tapped. Starting from the working
hypothesis that granular media are transiently elastic
while being deformed, we aim to understand the
notoriously complex plastic motion by combining two
simple and transparent elements, elasticity and stress
relaxation. In a recently published Letter~\cite{JL3},
we proposed a model for granular solids based on this
hypothesis. In the present manuscript, we give this
model a consistent, hydrodynamic framework, compatible
with conservation laws and thermodynamics.

The framework is valid for any healthy energy, but is
essentially devoid of predictive power if the energy
is left unspecified. Therefore, an explicit expression
for the total, conserved energy is given.
Encapsulating the key features of static granular
media: stress distribution, incremental stress-strain
relation, minimal and maximal density, the virgin
consolidation line, the Coulomb yield line and the cap
model, this expression should prove realistic enough
for rendering the specific hydrodynamic theory useful.
Much future work is needed to see whether further
agreement between theory and experiments may be
achieved, especially concerning cyclic loading,
tapping and shear band.

\appendix
\section{Equilibrium Conditions\label{appA}}
First, noting $\pi_{ij}{\rm d}u_{ij}=\pi_{ij}{\rm
d}\nabla_jU_i$ because $\pi_{ij}$ is symmetric, we write
the energy density per unit volume (dropping the
subscript of $w_0$ in this section) as
\begin{equation}\label{app1}
{\rm d}w=T{\rm d}s+\mu{\rm d}\rho-\pi_{ij}{\rm
d}\nabla_jU_i.
\end{equation}
Next, we vary the energy $\int w{\rm d}V$ for given
entropy $\int s{\rm d}V$, mass $\int \rho{\rm d}V$, and
for fixed displacement at the medium's surface, $\delta
U_i=0$. Taking $\ell_1, \ell_2$ as constant Lagrange
parameters and denoting the surface element as ${\rm
d}A_i$, we require the variation of the energy to be
extremal,
\begin{equation}\label{app2}
\delta\int \left(w-\ell_1s-\ell_2\rho\right)\,{\rm d}V =
0.
\end{equation}
Inserting Eq~(\ref{app1}) into (\ref{app2}), we find
\begin{eqnarray*}
\int\left[{T}\delta s+\mu\delta
\rho+{\pi_{ij}}\delta\nabla_jU_i-\ell_1\delta
s-\ell_2\delta\rho\right]\,{\rm d}V=
\\
\int\left[\left({T}-\ell_1\right)\delta
s+\left(\mu-\ell_2\right)\delta\rho
-\left(\nabla_j{\pi_{ij}}\right) \delta U_i\right]\,{\rm
d}V\\ \qquad+\oint {\pi_{ij}}\delta U_i\,{\rm d}A_i=0,
\end{eqnarray*}
where the last term vanishes because $\delta U_i\equiv0$
at the surface. If $\delta s$, $\delta\rho$ and $\delta
U_j$ vary independently, all three brackets must vanish.
And because $\ell_1, \ell_2$ are constant, $T,\mu$ also
need to be. So the conditions for the energy (or entropy)
being extremal are
\begin{equation}\label{app3}
\nabla_iT=0,\quad\nabla_i\mu=0,\quad
\nabla_j\pi_{ij}=0.
\end{equation}

In granular media for $\bar T_g=0$, density and
compression are coupled as
\begin{equation}\label{ap4}
{\rm d}u_{\ell\ell}=-{\rm d}\rho/\rho=\rho{\rm d}v,
\end{equation}
and do not vary independently. Simply inserting this
relation into the above calculation, we find
$\nabla_i(\mu+\pi_{\ell\ell}/3\rho)=0$ to replace the
last two conditions of Eq~(\ref{app3}). This is not the
correct result, because we have been varying the energy
and its variables above, keeping the volume unchanged
throughout, with $\delta U_i\equiv0$ at the surface. But
then $u_{\ell\ell}$ is fixed and cannot change with the
density $\rho$: Consider a one-dimensional medium between
$x=0$ and $x=x_0$, with $U_x(0), U_x(x_0)$ given. Since
$\nabla_j\pi_{ij}\sim
\partial^2_x[U_x(x_0)-U_x(0)]=0$, the one-dimensional strain
is $u_{\ell\ell}=\partial_x U_x=(U_x(x_0)-U_x(0))/x_0$
and cannot change.

To find the proper expression, we may take mass $M$
rather than volume $V$ as the constant quantity, and vary
the density by changing the volume, or the length in the
one-dimensional case. Holding $\delta U_i\equiv0$ at the
moving surface will then allow Eq~(\ref{ap4}) to hold.
Denoting the energy, entropy and volume per unit mass,
respectively, as $e\equiv w/\rho$, $\sigma\equiv s/\rho$,
$v\equiv V/M=1/\rho$, and $f\equiv w-Ts$, the equivalent
expression,
\begin{eqnarray}\label{app31}
{\rm d}e&=&T{\rm d}\sigma-P_T{\rm d}v-(\pi_{ij}/\rho){\rm
d}\nabla_jU_i,\\ P_T&\equiv&-w+Ts+\mu\rho=-f+\mu\rho,
\label{app32}
\end{eqnarray}
holds. Now we have $E=\int e{\rm d}M$, $S=\int\sigma{\rm
d}M$, $V=\int v{\rm d}M$, where ${\rm d}M=\rho{\rm d}V$
is the integrating mass element. Varying the energy for
given entropy, volume and requiring it to vanish, $\delta
E-\ell_1S-\ell_2V=0$, we find
\begin{equation*}
\int\left[\left({T}-\ell_1\right)\delta
\sigma+\left(P_T-\ell_2\right)\delta v\right]\,{\rm d}M
=\int\left(\nabla_j{\pi_{ij}}\right) \delta U_i\,{\rm
d}V.
\end{equation*}
implying $\nabla_iT=0$, $\nabla_i P_T=0$, and
$\nabla_j\pi_{ij}=0$. These are the same conditions as
Eq~(\ref{app3}), because $\nabla_i
P_T=s\nabla_iT+\rho\nabla_i\mu$. But if Eq~(\ref{ap4}) is
implemented, turning Eq~(\ref{app31}) to
\begin{eqnarray}\label{app33}
{\rm d}e&=&T{\rm d}\sigma-(P_T+\pi_{\ell\ell}/3){\rm
d}v-(\pi^0_{ij}/\rho){\rm d}\nabla_jU_i,\\ &=& T{\rm
d}\sigma-\rho^{-1}(P_T\,\delta_{ij}+\pi_{ij})\,{\rm
d}\nabla_jU_i,\end{eqnarray}
the equilibrium conditions are altered to become
\begin{equation}\label{app34}
\nabla_iT=0,\quad \nabla_i(P_T+\pi_{\ell\ell}/3)=0,\quad
\nabla_j\pi^0_{ij}=0.
\end{equation}
Clearly, the Cauchy, or total, stress in equilibrium is
given as
\begin{equation}
\sigma_{ij}=P_T\delta_{ij}+\pi_{ij},\,\, \text{with}\,\,
\nabla_j\sigma_{ij}=0.
\end{equation}

Including the gravitational energy $\rho\phi$ in $w$,
with $-\nabla_i\phi=G_i$, the gravitational constant
pointing downward, we have
\begin{equation}\label{app4}
{\rm d}w=T{\rm d}s+ (\mu+\phi){\rm d}\rho-\pi_{ij}{\rm
d}\nabla_jU_i+\rho\,{\rm d}\phi,\end{equation}
and find (via the same calculation as above) that
$\mu+\phi$ is now a constant, implying an alteration of
the second of Eqs~(\ref{app3}) to
\begin{equation}\label{app5}
\nabla_i\mu=G_i,
\end{equation}
or equivalently, $\nabla_i
P_T=s\nabla_iT+\rho\nabla_i\mu=\rho G_i$. If
Eq~(\ref{ap4}) holds, $\nabla_j\sigma_{ij}=0$ is
analogously changed to
\begin{equation}\label{app6}
\nabla_j\sigma_{ij}=\nabla_j(P_T\delta_{ij}+\pi_{ij})=\rho
G_i.
\end{equation}

\end{document}